\documentclass[12pt]{iopart}

\newcommand{\commentout}[1]{}

\usepackage{siunitx}
\usepackage{graphicx}
\usepackage{support-caption}
\usepackage[labelfont=bf, textfont=it]{caption}
\usepackage[labelfont=bf, textfont=it]{subcaption}
\usepackage[backend=biber, style=numeric-comp, sorting=none, maxcitenames=2, uniquelist=false]{biblatex}
\usepackage{hyperref}
\usepackage{amssymb}
\usepackage{ulem}

\usepackage{perpage} %the perpage package
\MakePerPage{footnote} %the perpage package command

\bibliography{MTM_ExB}

\begin{document}

\title[$\mathrm{E\times B}$ shear suppression of microtearing based transport in spherical tokamaks]{$\mathrm{E\times B}$ shear suppression of microtearing based transport in spherical tokamaks}

\author{B.S. Patel$^{1}$, M.R. Hardman$^2$\footnote{Current address: Tokamak Energy Ltd, 173 Brook Drive, Milton Park, Abingdon, OX14 4SD}, D. Kennedy$^1$, M. Giacomin$^{3, 4}$, D. Dickinson$^4$ and C.M. Roach$^1$}
%\email{bhav.patel@ukaea.uk}
\address{$^1$ UKAEA (United Kingdom Atomic Energy Authority), Culham Campus, Abingdon, Oxfordshire, OX14 3DB, United Kingdom}
\address{$^2$ Rudolf Peierls Centre for Theoretical Physics, University of Oxford, Oxford, OX1 3PU, UK}
\address{$^3$ Dipartimento di Fisica ``G. Galilei'', Università degli Studi di Padova, Padova, Italy}
\address{$^4$ University of York, Heslington, York, YO10 5DD, UK}

%
%\author{B. Patel}

%\address{York Plasma Institute, University of York, YO10 5DD}
\ead{bhavin.s.patel@ukaea.uk}

\vspace{10pt}
\begin{indented}
\item[]August 2023
\end{indented}

\begin{abstract}
Electromagnetic microtearing modes (MTMs) have been observed in many different spherical tokamak regimes. Understanding how these and other electromagnetic modes nonlinearly saturate is likely critical in understanding the confinement of a high $\beta$ spherical tokamak (ST). Equilibrium $\mathrm{E\times B}$ sheared flows have sometimes been found to significantly suppress low $\beta$ ion scale transport in both gyrokinetic simulations and in experiment. This work aims to understand the conditions under which $\mathrm{E\times B}$ sheared flow impacts on the saturation of MTM simulations, as there have been examples where it does [W. Guttenfelder \textit{et al} (2012)] and does not [H. Doerk \textit{et al} (2012)] have a considerable effect. Two experimental regimes are examined from MAST and NSTX, on surfaces that have unstable MTMs. The MTM driven transport on a local flux surface in MAST is shown to be more resilient to suppression via $\mathrm{E\times B}$ shear, compared to the case from NSTX where the MTM transport is found to be significantly suppressed. This difference in the response to flow shear is explained through the impact of magnetic shear, $\hat{s}$, on the MTM linear growth rate dependence on ballooning angle, $\theta_0$. At low $\hat{s}$, the growth rate depends weakly on $\theta_0$, but at higher $\hat{s}$, the MTM growth rate peaks at $\theta_0=0$, with regions of stability at higher $\theta_0$. Equilibrium $\mathrm{E\times B}$ sheared flows act to advect the $\theta_0$ of a mode in time, providing a mechanism which suppresses the transport from these modes when they become stable. The dependence of $\gamma^\textrm{MTM}$ on $\theta_0$ is in qualitative agreement with a recent theory [M.R. Hardman \textit{et al} (2023)] at low $\beta$ when $q\sim 1$, but the agreement worsens at higher $q$ where the theory breaks down. At higher $\hat{s}$, MTMs drive more stochastic transport due a stronger overlap of magnetic islands centred on neighbouring rational surfaces, but equilibrium $\mathrm{E\times B}$ shear acts to mitigate this. This is especially critical towards the plasma edge where $\hat{s}$ can be larger and where the total stored energy in the plasma is more sensitive to the local gradients. This work highlights the important role of the safety factor profile in determining the impact of equilibrium $\mathrm{E\times B}$ shear on the saturation level of MTM turbulence.

\end{abstract}

%\maketitle

\section{Introduction} \label{sec:intro}
Microtearing modes (MTMs) have been observed in gyrokinetic simulations of various conceptual spherical tokamak (ST) designs \cite{patel2022linear, kennedy2023electromagnetic, giacomin2024electromagnetic, wilson2020step} and in existing experiments in both the core \cite{doerk2012gyrokinetic, applegate2004microstability,valovic2011collisionality, guttenfelder2012scaling} and the pedestal \cite{dickinson2013microtearing, hatch2021microtearing}. These electromagnetic modes predominantly drive electron heat transport and can be destabilised by electron collisions \cite{drake1977kinetic}, which has been proposed as a candidate explanation for the $B\tau_E\propto \nu_{ee}^{-0.82}$ scaling seen in STs \cite{valovic2011collisionality,kaye2013NF}, with support from nonlinear gyrokinetic simulations of MTM turbulence \cite{guttenfelder2012simulation}. It is computationally challenging to achieve well converged saturated nonlinear simulations of MTM turbulence, but several such simulations suggest MTMs may play a significant transport role in spherical tokamaks \cite{guttenfelder2011electromagnetic,giacomin2023nonlinear}, and close to the edge in conventional aspect ratio devices when in H-mode \cite{hatch2021microtearing,doerk2011gyrokinetic}. To develop much needed reduced transport models for MTM turbulence with predictive power, it is important to understand the saturation mechanisms. There have been limited studies using simulations, and here we seek to explain the impact of flow shear on MTM turbulence. 
%Predictive modelling of plasmas where the transport involves MTMs requires the saturation of MTM turbulence using high fidelity first principle based models, which then must be captured in reduced core transport models. Whilst there have been recent advances in MTM simulations \cite{hatch2021microtearing, giacomin2023nonlinear} which are both saturated and converged, they have been historically elusive, therefore identifying potential mechanisms that can allow for MTM saturation is crucial.

For instabilities such as the ion temperature gradient (ITG) mode and kinetic ballooning mode (KBM), $\mathrm{E\times B}$ shear can reduce the turbulent transport \cite{burrell1997effects, davies2022kinetic}. This work aims to understand when $\mathrm{E\times B}$ shear is relevant in suppressing MTM transport. $\mathrm{E\times B}$ shear decorrelates turbulent eddies by tilting and shearing them radially, effectively adding a time dependence to their radial wavenumber $k_x$. One method to estimate the impact of flow shear on a mode is based on the dependence of its linear growth rate on the mode's radial wavenumber at the outboard mid-plane, $k_{x0}$, which is often parameterised using the ballooning angle $\theta_0 = k_{x0}/(k_y \hat{s})$ \footnote{For a circular, high aspect ratio, low $\beta$ un-shifted flux surface, $\theta_0$ corresponds to the poloidal angle at which the mode has zero radial wavenumber.}. Here $k_y= nq/r$ is the bi-normal wavenumber and $\hat{s}$ is the magnetic shear. At finite $\mathrm{E\times B}$ shear, modes at different $\theta_0$ become coupled, and the effective time average growth rate of a mode becomes an average of $\gamma^\mathrm{MTM}(\theta_0)$. The stabilising impact of $\mathrm{E\times B}$ shear therefore is stronger when the peak in $\gamma^\mathrm{MTM}(\theta_0)$ is narrower and more localised. The focus of this paper is to improve our understanding of the factors determining $\gamma^\mathrm{MTM}(\theta_0)$ and the corresponding susceptibility of MTM turbulence to suppression through $\mathrm{E\times B}$ shear.

%$\mathrm{E\times B}$ shear predominantly acts in two ways. Firstly $\mathrm{E\times B}$ shear acts to decorrelate turbulent eddies by ``tilting" them, effectively adding a time dependence to their radial wavenumber $k_x$. This couples unstable low $k_x$ modes to stable higher $k_x$ modes. One method of determining the impact of this mechanism is to look at a modes linear stability properties as its $k_x$ is increased at the outboard mid-plane. This can be parameterised using the ballooning angle $\theta_0 = \frac{k_x}{k_y \hat{s}}$ which, for a circular un-shifted flux surface, corresponds to the poloidal angle at which the mode has zero radial wavenumber. Here $k_y= \frac{nq}{r}$ is the bi-normal wavenumber and $\hat{s}$ is the magnetic shear. Throughout this work we will perform linear gyrokinetic scans in $\theta_0$ to determine if the linear mode is stabilised at non-zero $\theta_0$ and thus if the transport can suppressed by making $\theta_0 = \theta_0(t)$. Throughout this work we will perform linear gyrokinetic scans in $\theta_0$ to determine if the linear mode is stabilised at non-zero $\theta_0$ and thus if the transport can suppressed by making $\theta_0 = \theta_0(t)$.

MTMs can of course saturate via other mechanisms such as zonal fields \cite{pueschel2013properties, giacomin2023nonlinear}, local electron temperature gradient flattening \cite{ajay2022, hatch2021microtearing} and coupling to dissipative modes \cite{doerk2011gyrokinetic}, though that will not be a particular focus here. %\sout{as they are outside of experimental control, whereas $\mathrm{E\times B}$ shear can be directly influenced via momentum injection.}

This paper also examines the applicability of recent work done by Hardman \textit{et al} \cite{hardman2023new}, where a theory is derived for electromagnetic electron-driven instabilities resembling MTMs, that have current layers localised to mode-rational surfaces and bi-normal wavelengths comparable to the ion gyroradius. The gyrokinetic equation is derived for two different regions, one inner region localised around the rational surface. Secondly an outer region far away from the rational surface at the centre of the flux tube in the local gyrokinetics simulation. In ballooning space the inner region corresponds to $\theta \gg 1$, and the outer region corresponds to $\theta \lesssim 1$. In this theory a mass ratio expansion is taken with the following ordering for $\beta$

\begin{equation}
    \beta \sim \bigg(\frac{m_e}{m_i}\bigg)^{\frac{1}{2}} \sim k_y\rho_e \ll 1
\end{equation}

\noindent and an asymptotic matching condition is applied to solutions from the two regions to obtain the dispersion relation. This theory exposes an important local equilibrium parameter, $\beta_\mathrm{eff}$, that increases the MTM instability drive when it is large. $\beta_\mathrm{eff}$ is defined as:

\begin{equation}
    \label{eqn:beta_eff}
    \beta_{\textrm{eff}} = \beta_e \frac{2\pi G(\theta_0)}{\hat{s}k_y\rho_e}
\end{equation}

\noindent where 

\begin{equation}
    G(\theta_0) = \frac{1}{qR_0}\frac{\hat{s}}{\pi} \int_{-\infty}^{\infty} \frac{B^2}{B_\textrm{ref}} \frac{k_y^2}{k_\perp^2} \frac{d\theta}{B.\nabla\theta}
\end{equation}

\noindent with all the $\theta_0$ dependence of $\beta_\textrm{eff}$ being contained in $k_\perp$. $G(\theta_0)$ is highly sensitive to how $k_\perp$ varies along the field line and it is generally maximised when $\theta_0=0$. The integrand's $\theta_0$ dependence is dominated by the factor $k_y^2/k_\perp^2$, and it is maximised at $k_\perp=k_y$, i.e. at $k_x=0$. In a large aspect ratio circular geometry, $k_y^2/k_\perp^2 = (1+\hat{s}^2(\theta - \theta_0)^2)^{-1}$, which is largest either when $\theta = \theta_0$ or when $\hat{s}$ is low. In such geometries $G(\theta_0)= 1 + \mathcal{O}(r/R)$.

A more physical picture for $G(\theta_0$) can be built by examining the linearised form of Amp\`ere's Law for the perturbed current and perpendicular magnetic field:
\begin{equation}
    k_\perp^2 A_{||} = \frac{4\pi}{c} J_{||} \label{eqn:amp}
\end{equation}
MTMs generate re-connection whereby equilibrium field lines undergo a finite radial displacement over their trajectory from $\theta = -\infty \rightarrow \infty$. The radial displacement of a field line is given by:
\begin{equation}
    \label{eqn:del_psi}
    \Delta \Psi = \int_{-\infty}^{\infty} \frac{ k_y A_{||} d\theta}{b.\nabla \theta} = \frac{4\pi}{c} \int_{-\infty}^{\infty} \frac{J_{||}}{B} \frac{B^2 k_y}{k_\perp^2}\frac{d\theta}{B\cdot\nabla\theta}
\end{equation}
Quasi-neutrality requires a divergence-free perturbed current, $\mathbf{\nabla}.(J_{\parallel}\mathbf{b})=0$\, resulting in:
\begin{equation}
    \label{eqn:j_par}
    B\cdot\nabla\theta\frac{\partial}{\partial \theta} \bigg(\frac{J_{||}}{B}\bigg) = 0
\end{equation}

Here the perpendicular current $J_\perp$ has been ordered out by the low $\beta$ assumption which will ignore the ion contribution to the current. Combining equations \ref{eqn:del_psi} and \ref{eqn:j_par} whilst dropping constants gives:
\begin{equation}
    \Delta \Psi \propto \int_{-\infty}^{\infty} \frac{B^2}{k_\perp^2} \frac{d\theta}{B\cdot\nabla\theta} \label{eqn:dpsi}
\end{equation}
\noindent where the integrand is proportional to $G(\theta_0)$. This exposes how $G(\theta_0)$, and thus $\beta_\mathrm{eff}$, represent a local equilibrium geometry parameter to which the radial field line displacement is proportional for a given perturbed parallel current $J_{\parallel}/B$. $\beta_\mathrm{eff}$ determines how efficiently a magnetic field perturbation can tap energy from the current perturbation generated by the electron temperature gradient drive, which is key in setting the MTM growth rate.

In this paper we will explore the crucial dependence of the growth rate on $\theta_0$, which has received relatively little attention in the literature. Section \ref{sec:exp} outlines the local equilibria and grid parameters used for gyrokinetic calculations of MTMs that will be presented for MAST and NSTX plasmas. Section \ref{sec:mast} examines MTMs previously found in MAST \cite{valovic2011collisionality}, using both linear and nonlinear simulations. The impact of $\theta_0$ on these modes is determined and we assess whether $\beta_\mathrm{eff}$ is useful as an indicator of the linear instability drive. In Section \ref{sec:nstx}, a similar approach is applied to an NSTX plasma \cite{guttenfelder2012simulation, guttenfelder2011electromagnetic}, where the MTM turbulence is found to be much more susceptible to stabilisation via $\mathrm{E\times B}$ shear, as opposed to the MAST surface examined. This difference is explained by the impact of higher $\hat{s}$ of the NSTX plasma on the linear stability. This all points to the importance of tailoring the safety factor profile, which is important in determining when MTMs are susceptible to sheared flow stabilisation. %\sout{and having the ability to drive sheared flows if the transport caused by MTMs needs to be minimised.}

\section{Equilibria and numerical set up}\label{sec:exp}
Many different codes have been used to analyse MAST and NSTX plasmas, both linearly and nonlinearly. For the cases studied here, we use CGYRO \cite{candy2016crucial}.

The Miller representation \cite{miller1998noncircular, arbon2020rapidly} was used to describe the local equilibrium parameters of each chosen surface from MAST and NSTX, with parameters outlined in Table \ref{tab:miller_params}. Gradients are defined such that $a/L_X = -a/X\frac{\partial X}{\partial r}$ where $a$ is the minor radius of the last closed flux surface. The level of $\mathrm{E\times B}$ shear is parameterised by $\gamma_\mathrm{E\times B} = - r/q \frac{\partial \omega_0}{\partial r}$, with $\omega_0$ being the local toroidal angular rotation frequency of the plasma. All heat fluxes in this paper are normalised to $Q_{gB} = n_e T_e c_s (\rho_s / a)^2$ where $\rho_s = c_s(e B_{unit}/m_Dc)$, with $c_s=\sqrt{T_e/m_D}$ and $B_{unit} = q/r \frac{\partial \psi}{\partial r}$.

\setlength{\tabcolsep}{8pt}
\begin{table}[!htb]
    \centering
    \begin{tabular}{c c c c c c c c}
        & $r/a$ & $R_{\mathrm{maj}}/a$ & $\partial_r R_\mathrm{maj}$ &  $a/L_{n}$ & $a/L_{T,e}$ & $q$ &  $\hat{s}$ \\
        \hline\hline
        MAST & 0.51 & 1.57 & -0.13 & 0.22 & 2.11 & 1.08 & 0.34 \\
        NSTX & 0.60 & 1.53 & -0.29 & -0.83 & 2.73 & 1.71 & 1.70 \\
        \hline
        \\
        &  $\gamma_\mathrm{E\times B} (c_s/a)$  &$\nu_{ee} (c_s/a)$& $\kappa$ & $s_\kappa$ & $\delta$ & $s_\delta$ & $\beta'$ \\
        \hline\hline
        MAST & 0.19 & 0.82 & 1.41 & 0.01 & 0.16 & 0.12  & -0.53 \\
        NSTX & 0.18 & 1.45 & 1.71 & 0.11 & 0.13 & 0.17  & -0.36 \\
        \hline
        \\
        & $n_\mathrm{e} (\mathrm{m}^{-3})$ & $T_\mathrm{e} (\mathrm{keV}) $ & $a (\mathrm{m})$ & $B_0 (\mathrm{T})$ & $B_\mathrm{unit} (\mathrm{T})$ & $\beta_e$ & $\beta_{e,\mathrm{unit}}$     \\
        \hline\hline
        MAST & $3.55\times 10^{19}$ & 0.44 & 0.57 & 0.33 & 0.54 & 0.11 & 0.023 \\
        NSTX & $6.01\times 10^{19}$ & 0.45 & 0.62 & 0.32 & 0.66 & 0.06 & 0.025  \\
        \hline
    \end{tabular}
    \caption{Local Miller parameters and reference values for the equilibrium flux surfaces simulated in this work from MAST \#22769 and NSTX \#120968, where the parameters are defined as in \cite{miller1998noncircular}. Here $\beta_e$ is the electron plasma $\beta$ normalised to $B_0 = f / R_\mathrm{maj}$ and $\beta_{e,\mathrm{unit}}$ utilises $B_\mathrm{unit}$ as the normalising field.}
    \label{tab:miller_params}
\end{table}

This work examines ST scenarios where MTMs have been previously found. Firstly, the MAST discharge \#22769 at the flux surface with $r/a=0.51$ at $t=0.2s$, as discussed in \cite{valovic2011collisionality}. This surface along with an outer flux surface close to the peak in experimental collisionality is examined in detail in \cite{giacomin2023nonlinear}. Furthermore, the flux surface with $r/a=0.6$ in the NSTX discharge \#120968 at $t=0.56s$ is examined here, which has been discussed previously in \cite{guttenfelder2011electromagnetic}.
 
The aim of this study is to examine MTM in different regimes. Although the MTM is generally dominant in the local equilibria analysed here, different modes can become the dominant instability during parameter scans. The following choices were made to avoid mode transitions and maximise the likelihood of the MTM remaining the dominant instability. All simulations in this work were performed without $\delta B_{||}$ fluctuations and $a/L_{T,i}=0.0$. This reduces the linear drive for other instabilities, such as KBMs and ITGs, without affecting the MTM drive \cite{patel2021confinement}. This can artificially preserve MTM as the the dominant instability, making it easier to track the mode in isolation linearly. Note that the focus here is not to quantitatively predict the transport from the mode, but rather to determine the sensitivity of growth rates for particular modes to $\theta_0$.

Linear calculations were conducted using 64 $\theta$ grid points, 8 energy grid points and 24 pitch angle points with 64 connected $2\pi$ segments. For simplicity only Lorentz collisions were included in the simulations as this was found to be sufficient for unstable MTMs. $Z_\textrm{eff}= 1.0 $ was used and only 2 species were simulated, electrons and deuterium. An additional filter has been used to classify a mode as an MTM by imposing a threshold level of re-connection from the perturbed radial magnetic field by requiring the field line tearing parameter, $C_{\mathrm{tear}} > 0.1$ where:
\begin{equation}
    \centering
    C_{\mathrm{tear}} = \frac{|\int A_{||} dl|}{\int |A_{||}| dl}
\end{equation}
\textbf{Pyrokinetics}, a python library which aims to standardise gyrokinetic analysis \cite{Patel_pyrokinetics_2022}, was used to generate the input files and perform the analysis in this work.

\section{MAST \#22769}
\label{sec:mast}
Using the MAST local equilibrium parameters in Table \ref{tab:miller_params}, from \cite{valovic2011collisionality}, the micro-stability of this equilibrium was explored as a function of $k_y\rho_s$ and $\theta_0$, focusing on the ion scale in the bi-normal direction with simulations performed up to $k_y\rho_s=1.1$.

\subsection{Linear simulations} \label{sec:mast_lin}
\begin{figure}[!htb]
    \begin{subfigure}{0.49\textwidth}
        \centering
        \includegraphics[width=74mm]{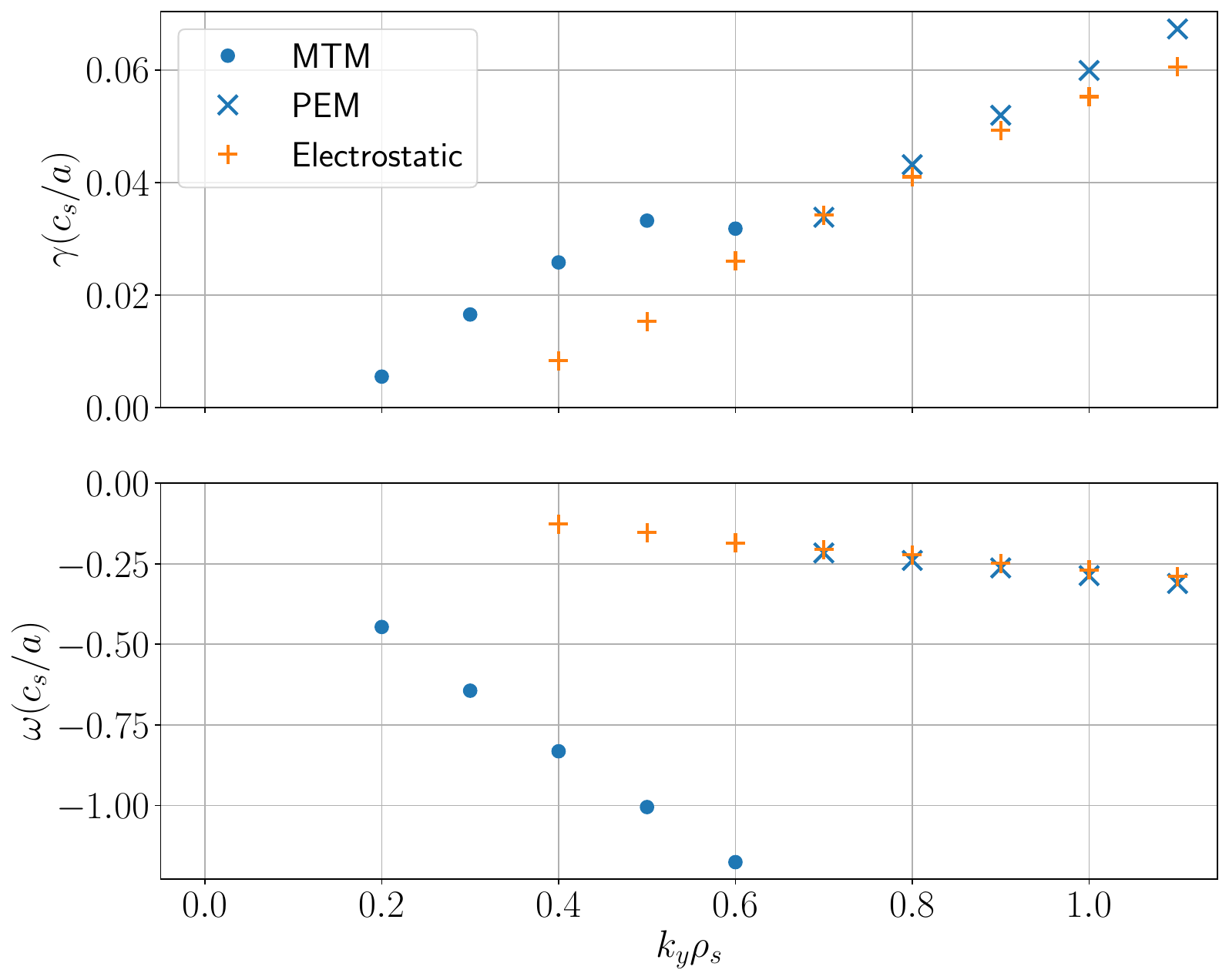}
        \caption{}
        \label{fig:mast_eigval}
    \end{subfigure}
    \begin{subfigure}{0.49\textwidth}
        \centering
        \includegraphics[width=77mm]{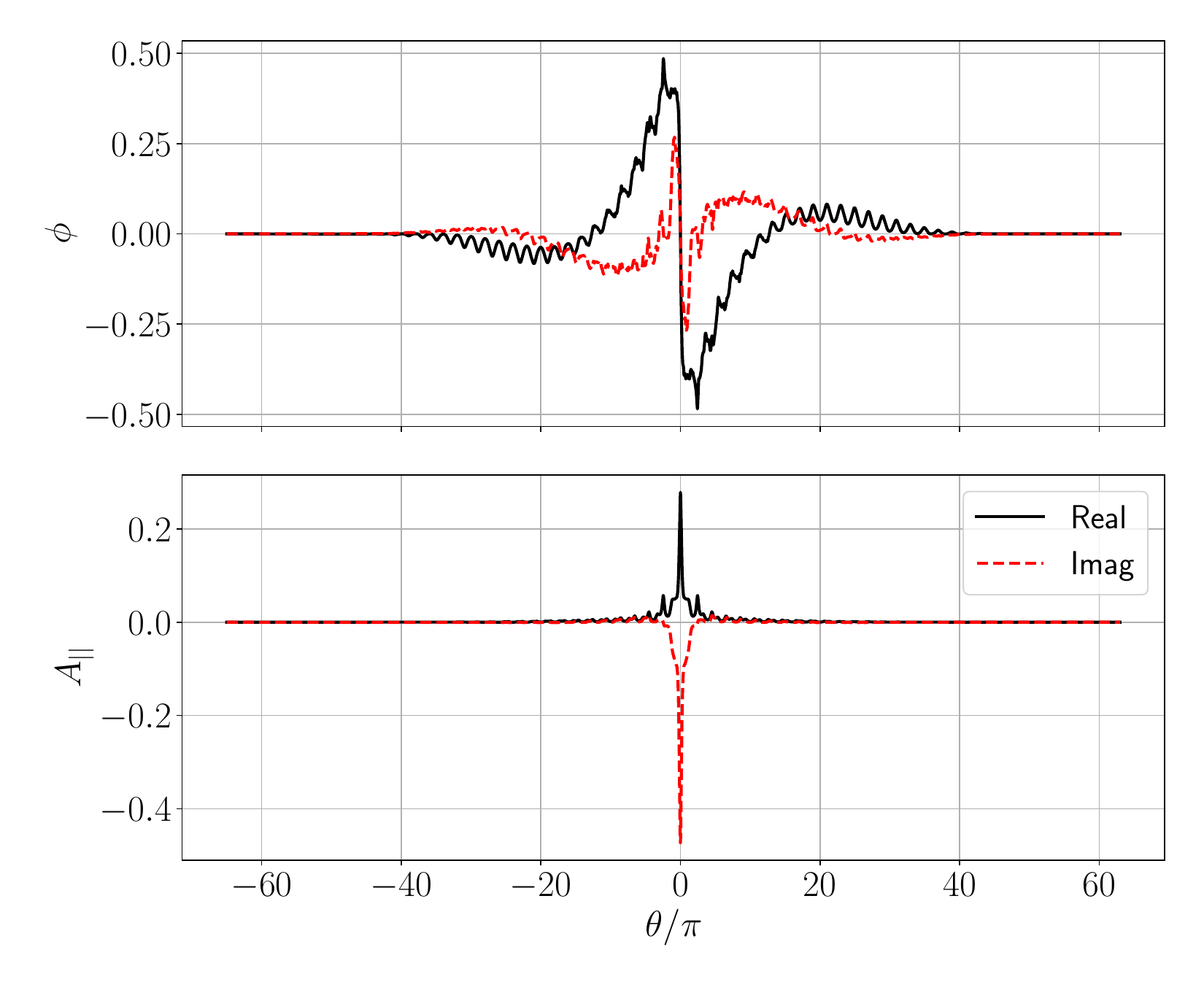}
        \caption{}
        \label{fig:mast_eigfunc_ky05}
    \end{subfigure}
    \caption{a) Eigenvalues for the MAST local equilibrium at $\theta_0=0$, with the dominant instability of an electromagnetic simulation shown in blue and an electrostatic simulation shown in orange. The dots illustrate a MTM and the crosses a PEM. b) $\phi$ and $A_{||}$ eigenfunctions of the MTM at $k_y\rho_s=0.5$.}
    \label{fig:mast_base}
\end{figure}

Figure \ref{fig:mast_eigval} shows in blue the growth rate, $\gamma$, and mode frequency, $\omega$, of the dominant linear instabilities at $\theta_0=0$. For $k_y\rho_s\leq 0.6$, the dominant mode is an MTM and the eigenfunction of the most unstable MTM at $k_y\rho_s=0.5$ is illustrated in Figure \ref{fig:mast_eigfunc_ky05}. This exhibits the conventional properties of an MTM in that $\phi$ has odd parity whilst $A_{||}$ has even parity. $\phi$ is significantly extended in ballooning space whilst $A_{||}$ is very narrow. This MTM is ion scale in the bi-normal direction, but these eigenfunctions illustrate its multi-scale nature in the radial direction: low $k_x$ is needed to resolve both $\phi$ and $A_{||}$ in the outer layer, but very high $k_x$ is also required to resolve $\phi$ in the inner layer.

An electrostatic mode becomes dominant at $k_y\rho_s>0.6$, where the MTM becomes sub-dominant\footnote{This differs from results reported in \cite{giacomin2023nonlinear} for the same local equilibrium, where no overlap of modes was seen. This is due to a difference in the collision operators used. \cite{giacomin2023nonlinear}used a Sugama operator with more physics, whilst for simplicity a Lorentz operator was chosen here.}, and this is confirmed through an electrostatic simulation without $A_{||}$, shown in orange, where the mode frequency and growth rate are largely unchanged on removal of $A_{||}$. Furthermore, this mode is clearly unstable when $0.4 \leq k_y \rho_s \leq 0.6$, but is sub-dominant to the MTM. This mode has a frequency in the electron diamagnetic direction. It has an even parity $\phi$ eigenfunction and the linear fluxes indicate that it drives predominantly electron heat transport, with little ion and particle transport. It has a similar transport signature to MTM, though is definitely not an MTM given its predominantly electrostatic nature and the fact that $C_\mathrm{tear}\approx 0$. Its electrostatic potential eigenfunction looks very similar to those found in \cite{hallatschek2005giant, hardman2023new, hardman2022extended, patel2022linear} for a radially localised ETG mode, and in this work is denoted as electrostatic passing electron mode (PEM).

\begin{figure}[!htb]
    \centering
    \includegraphics[width=80mm]{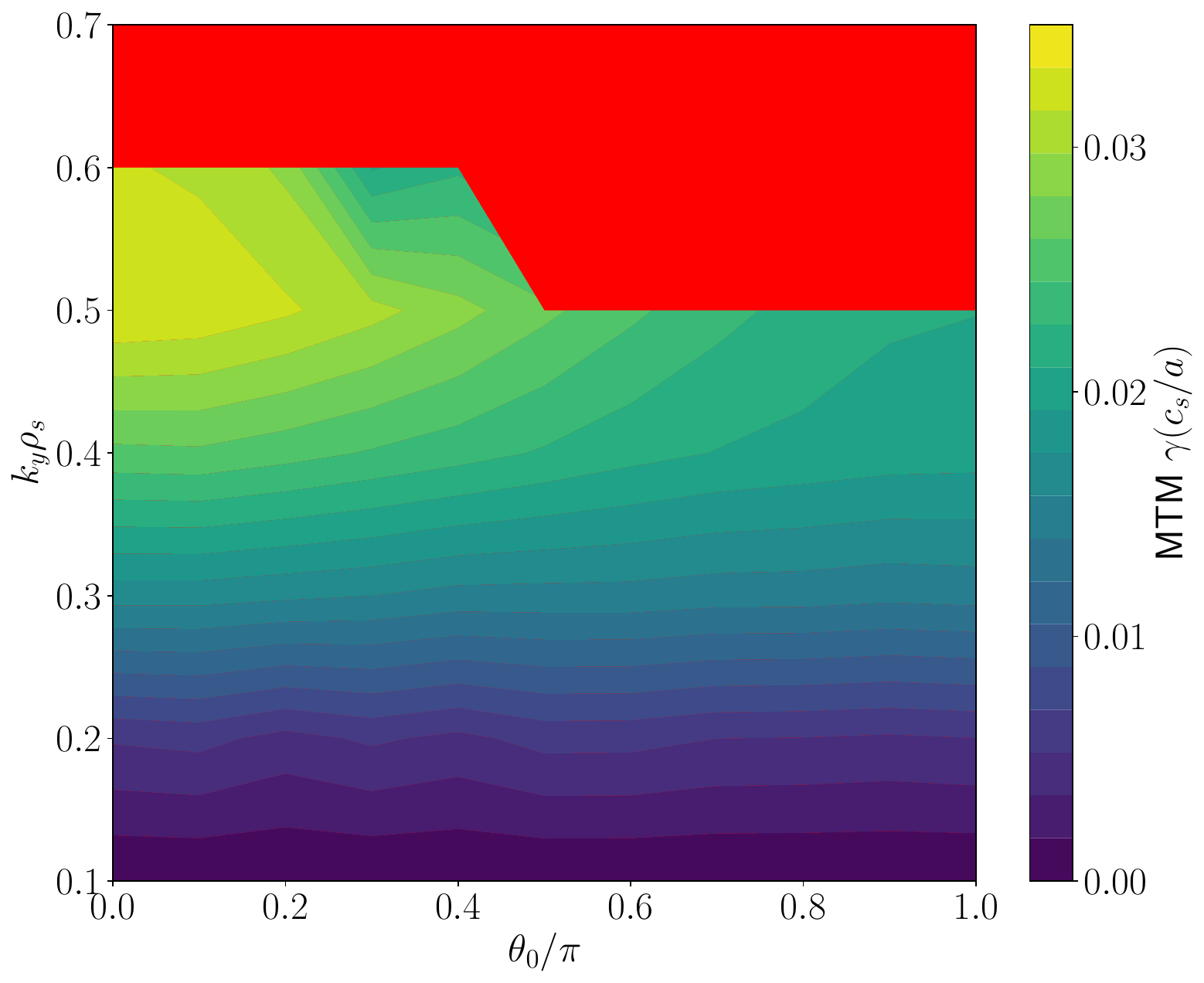}
    \caption{A 2D contour plot showing the growth rate $\gamma$ of MTMs (where they dominate) for the MAST local equilibrium, plotted against $\theta_0$ and $k_y\rho_s$. The blue-yellow contours denote the MTM growth rate, whilst the shaded red region shows where the PEM is dominant. The modes were differentiated using the field line tearing parameter $C_\textrm{tear}$.}
    \label{fig:theta0_ky_scan_mast_gam}
\end{figure}

Figure \ref{fig:theta0_ky_scan_mast_gam} illustrates a 2D scan in $k_y\rho_s$ and $\theta_0$ that was performed to see how $\gamma^\mathrm{MTM}$ varies with $\theta_0$, though this picture is somewhat complicated in the region $k_y \rho_s>0.5$ by the PEM. The blue-yellow contours indicate the MTM growth rate where it is dominant, whilst the shaded red region at higher $k_y \rho_s$ is where the PEM is dominant. At $k_y\rho_s \leq 0.5$, the MTM remains the dominant mode across $\theta_0$ and is slightly stabilised with increasing $\theta_0$. The dependence on $\theta_0$ is weaker as $k_y$ is reduced. This suggests that $\mathrm{E\times B}$ shear will have a limited impact on these modes nonlinearly. The PEM growth rate also weakly depends on $\theta_0$ for this surface.

\begin{figure}
    \begin{subfigure}{0.49\textwidth}
        \centering
        \includegraphics[width=75mm]{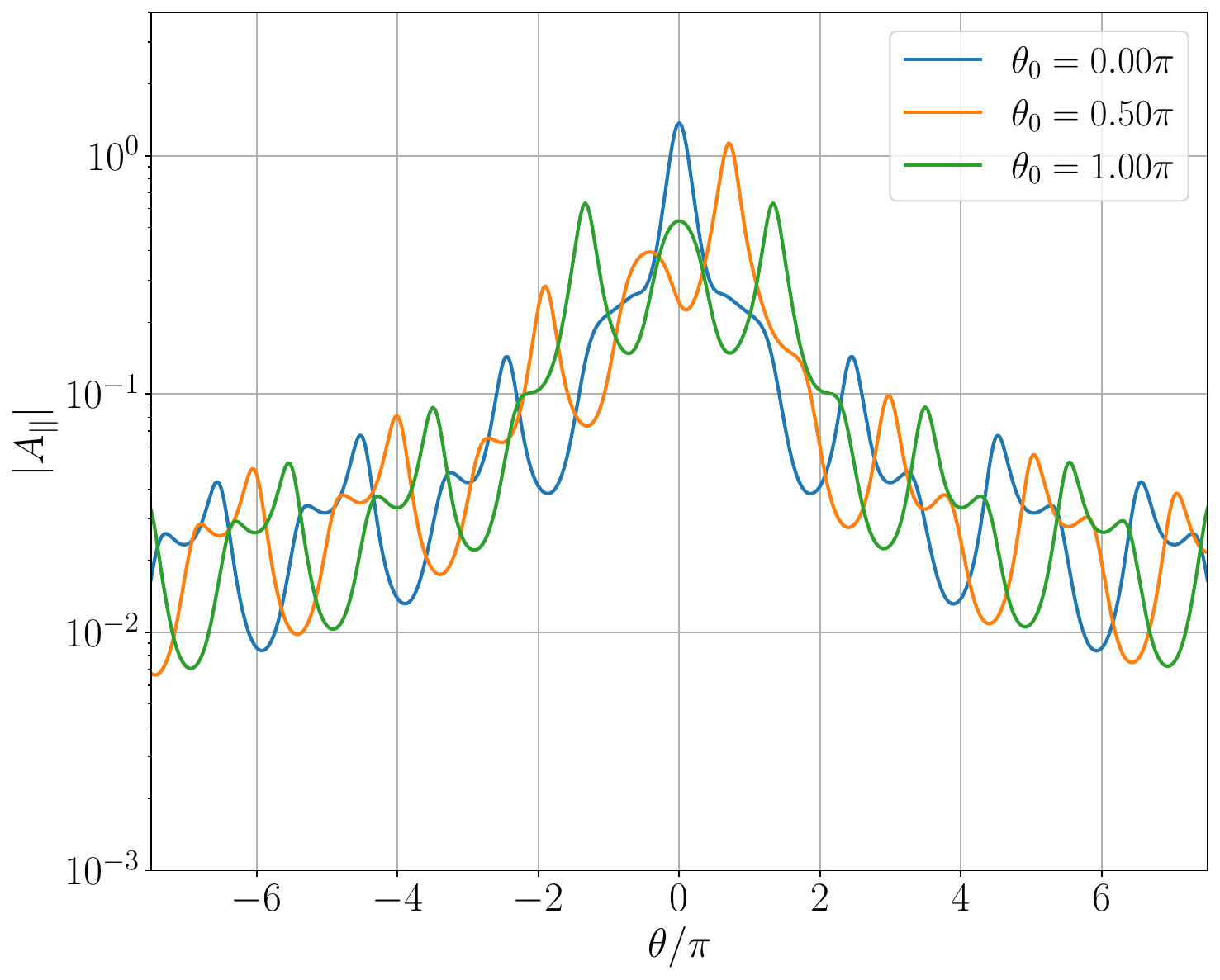}
        \caption{}
        \label{fig:theta0_scan_mast_eig}
    \end{subfigure}
    \begin{subfigure}{0.49\textwidth}
        \centering
        \includegraphics[width=75mm]{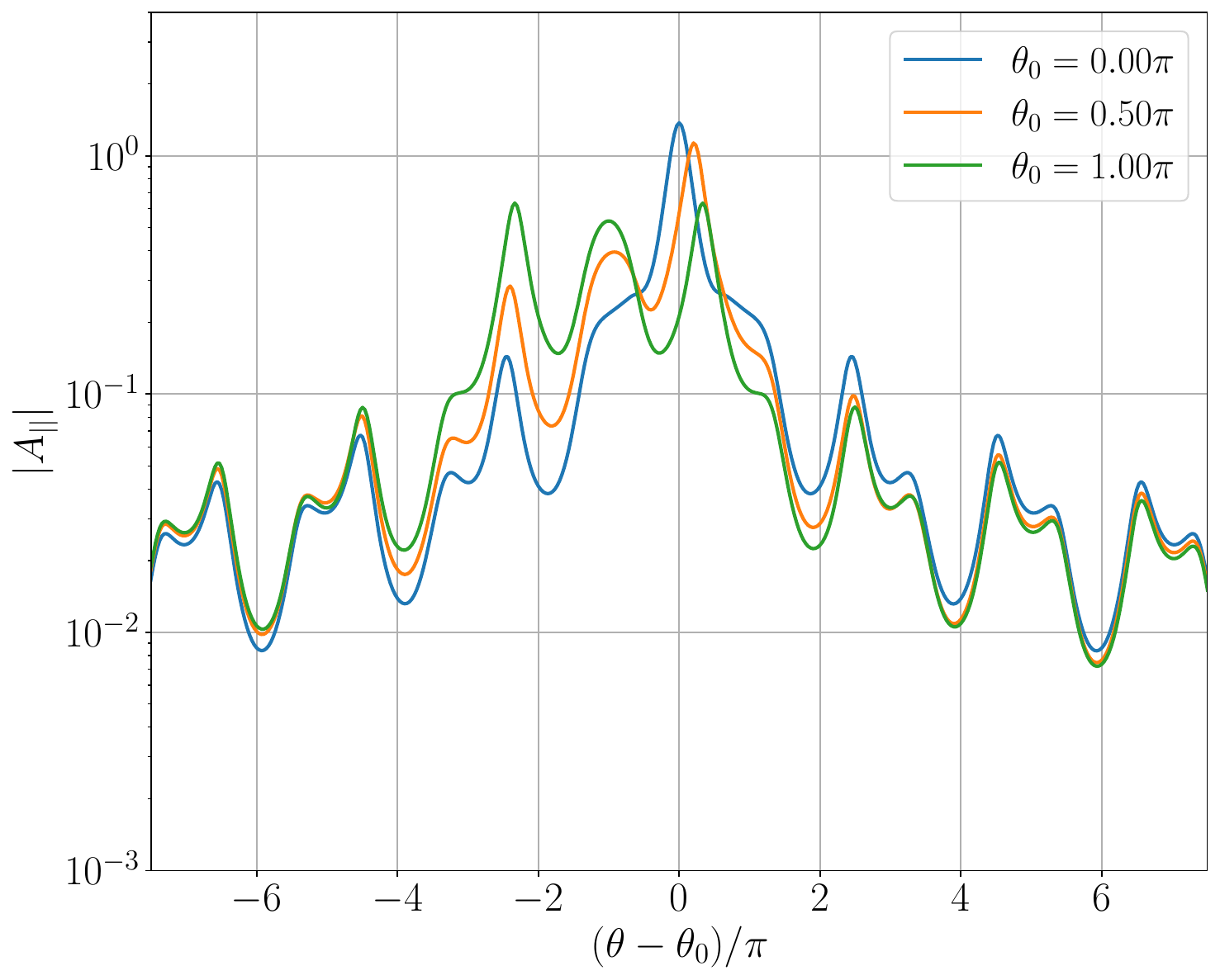}
        \caption{}
        \label{fig:theta0_scan_mast_eig_shift}
    \end{subfigure}
    \caption{The eigenfunction of the MTM in MAST at $k_y\rho_s=0.5$ at various different $\theta_0$. a) plots against the $\theta$, whilst b) shifts the eigenfunctions by $\theta_0$. Doing so lines up the eigenfunction when $|\theta| > \pi$.}
    \label{fig:psin_0.5_eigfunc}
\end{figure}

The $A_{||}$ eigenfunctions at $k_y\rho_s=0.5$ are shown for $\theta_0 = 0$, $0.5\pi$ and $\pi$ in Figure \ref{fig:theta0_scan_mast_eig}, which is the highest $k_y\rho_s$ where the MTM is the dominant instability throughout $\theta_0$. The peak of the eigenfunction moves away from $\theta = 0$ as $\theta_0$ increases up to $\pi$. The periodic behaviour seen in the tails of the eigenfunction also shift with $\theta_0$, such that when the axes are shifted by $\theta_0$, the peaks and troughs of the eigenfunctions line up perfectly, demonstrated in Figure \ref{fig:theta0_scan_mast_eig_shift}. This indicates that the location of the peaks in the tail of the distribution function is impacted by the location of the peak around $\theta \sim 0$ and the troughs occur where $(\theta-\theta_0) \bmod 2\pi = 0 $. 

\begin{figure}[!htb]
    \begin{subfigure}{0.49\textwidth}
        \centering 
        \includegraphics[width=75mm]{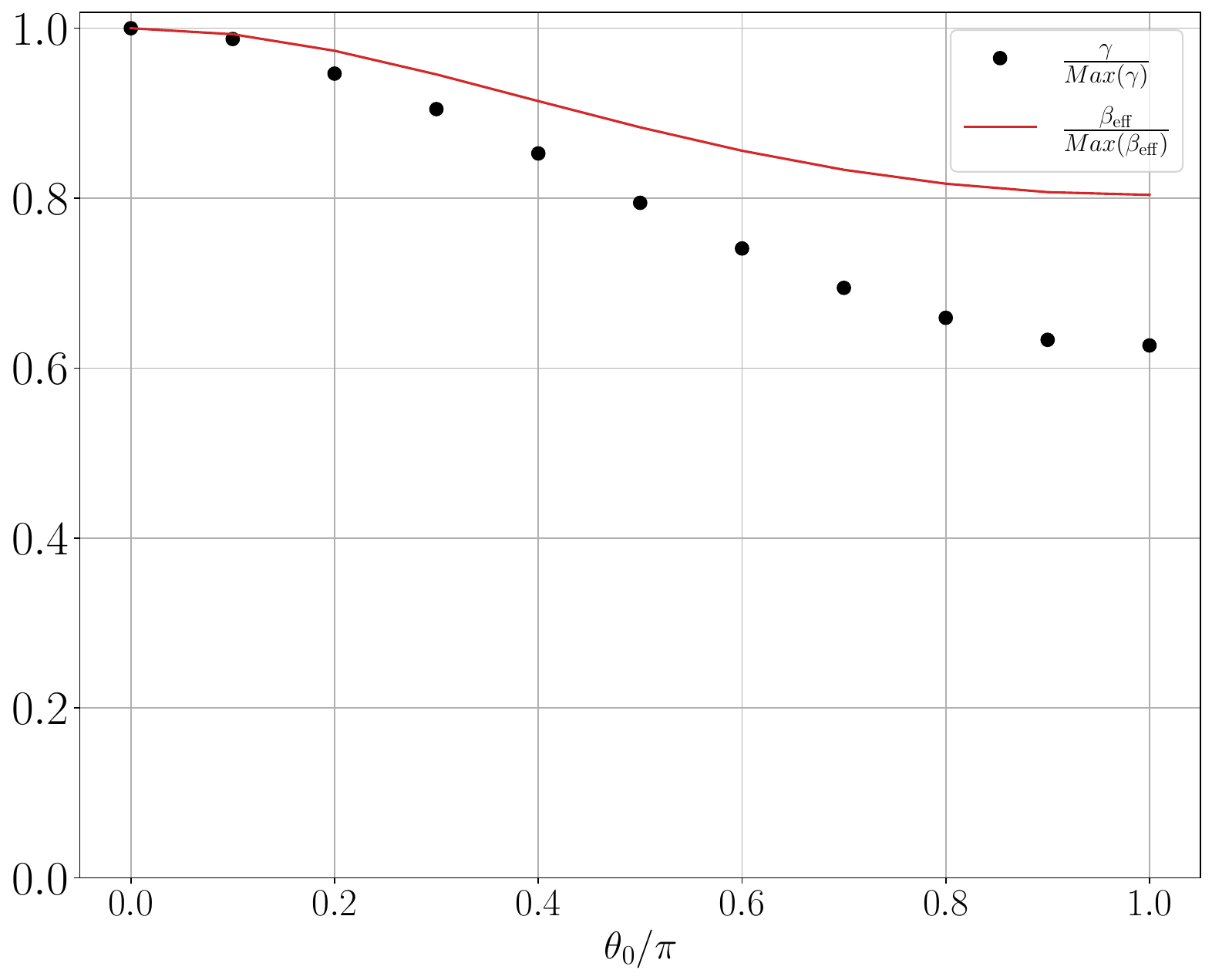}
        \caption{}
        \label{fig:mast_gam_betaeff}
    \end{subfigure}
    \begin{subfigure}{0.49\textwidth}
        \centering
        \includegraphics[width=75mm]{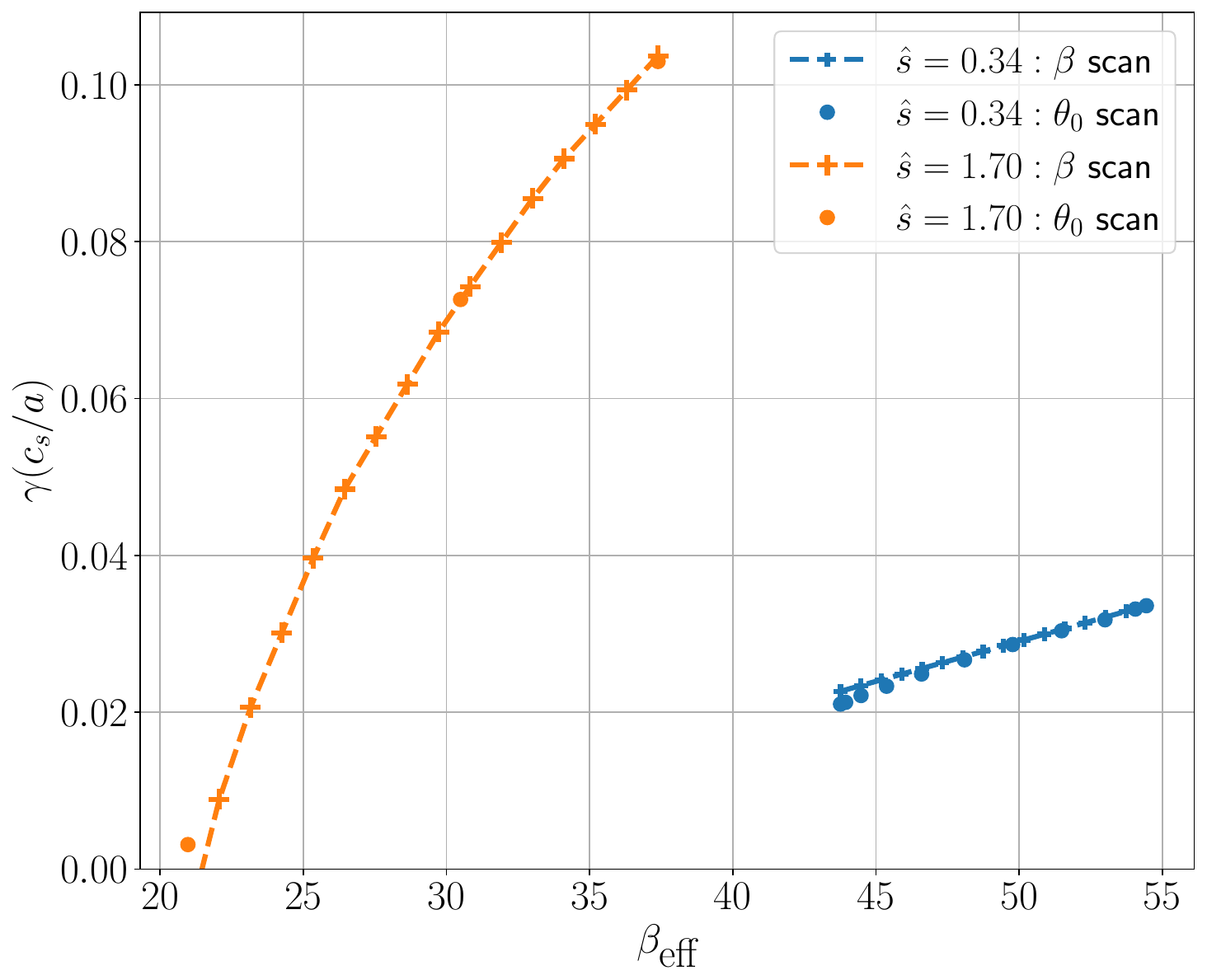}
        \caption{}
        \label{fig:mast_beta_eff_scan}
    \end{subfigure}
    \caption{a) The growth rate (black) and $\beta_\mathrm{eff}$ (red) normalised to their respective maxima for the MAST MTM at $k_y\rho_s=0.5$ against $\theta_0$. b) A $\beta_\mathrm{eff}$ scan via changing $\beta_e$ (dashed lines) and $\theta_0$ (dots). This was done at the equilibrium $\hat{s}=0.34$ (blue) and at the NSTX value of $\hat{s}=1.70$ (orange).}
    \label{fig:mast_beta_eff}
\end{figure}

It is interesting to assess whether the $\gamma^\mathrm{MTM}$ dependence on $\theta_0$, which decreases slightly with rising $\theta_0$ for $k_y\rho_s>0.4$, can be understood from a theoretical point of view.
%and we find that this behaviour is consistent with the $\theta_0$ dependence of $\beta_\mathrm{eff}$ . 
$\beta_\textrm{eff}$, defined in Section \ref{sec:intro}, is calculated for this MAST case and is compared with $\gamma^\mathrm{MTM}$ in Figure \ref{fig:mast_gam_betaeff}. Both monotonically decrease with $\theta_0$, in a similar trend, supporting the suggestion from the theory that the MTM driving mechanism is less effective at lower $\beta_\textrm{eff}$.

Equation \ref{eqn:beta_eff} shows that $\beta_\textrm{eff}$ is sensitive to $\theta_0$, $\beta_e$, $k_y$ and $\hat{s}$, so the behaviour of the MTM can be examined whilst modifying these parameters. Figure \ref{fig:mast_beta_eff_scan} shows $\gamma^\mathrm{MTM}$ against $\beta_\mathrm{eff}$ for two independent scans, firstly in $\theta_0$ (as shown in Figure \ref{fig:mast_gam_betaeff}) and secondly in $\beta_e$ (at fixed $\theta_0=0$) to scale $\beta_\mathrm{eff}$ over the same range from the $\theta_0$ scan. These scans were performed using the MAST local equilibrium parameters with two values of $\hat{s}$: the local equilibrium value of $\hat{s}=0.34$ shown in blue; and a higher value of $\hat{s}=1.70$, corresponding to the value of the NSTX equilibrium discussed in Section \ref{sec:nstx}, indicated by the orange markers. $\gamma^\mathrm{MTM}$ is found to be a unique function of $\beta_\mathrm{eff}$ for each local equilibrium. Note, the higher $\hat{s}$ case is more unstable at a lower $\beta_\mathrm{eff}$, indicating that although $\beta_\mathrm{eff}$ is an important parameter for MTM stability, it is not the only one.

If the critical threshold for unstable MTMs in $\beta_{e}$ at $\theta_0=0$ can be determined for a given local equilibrium, this will give the critical $\beta_\mathrm{eff}$ for the instability. Then from geometry alone, it should be possible to assess where $\beta_\mathrm{eff}$ drops below this critical value as $\theta_0$ increases, and thus determine whether the mode goes stable.

In this MAST equilibrium, the weak dependence of $\gamma^\mathrm{MTM}$ (and $\beta_\mathrm{eff}$) on $\theta_0$ suggests that MTMs should only be very weakly impacted by equilibrium $\mathrm{E\times B}$ shear.

\subsection{Nonlinear simulations}
\begin{figure}[!htb]
    \centering
    \includegraphics[width=80mm]{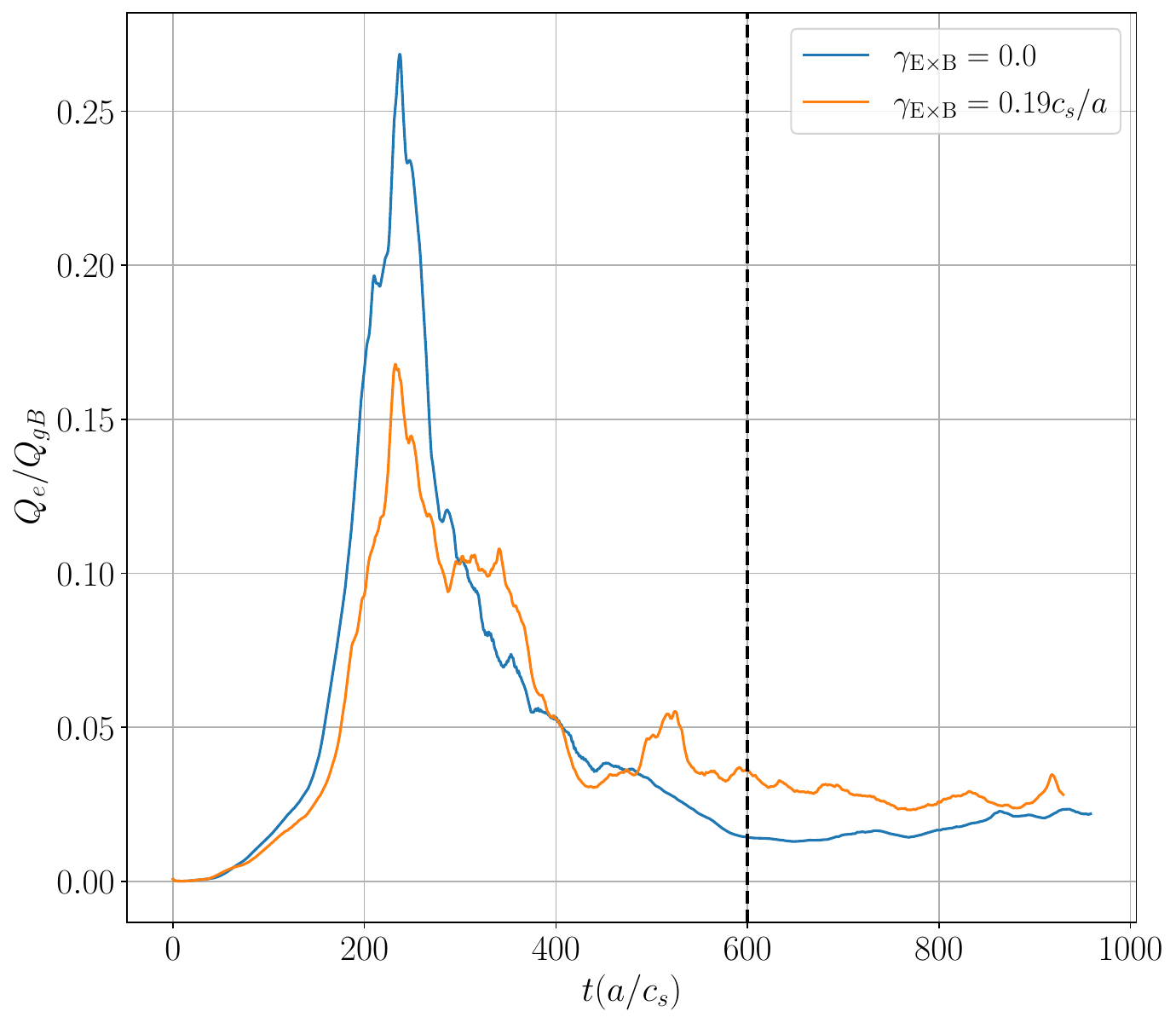}
    \caption{Nonlinear electron heat flux prediction for the MAST case when $\gamma_\mathrm{E\times B} = 0.0$ (blue) and $\gamma_\mathrm{E\times B} = 0.19 c_s/a$ (orange). Note that the electromagnetic electron heat flux dominates the total flux driving $>$82\% of the total heat transport in both of these simulations. The vertical dashed line denotes the time from which the average and uncertainty in the flux is calculated. The particle and the ion heat fluxes are two orders of magnitude smaller compared to the electron heat flux.}
    \label{fig:mast_nl_flux}
\end{figure}

Nonlinear simulations were performed to assess the impact of $\mathrm{E\times B}$ shear on MTM turbulence in this MAST equilibrium. These simulations required 256 $k_x$ grid points with a $k_{x,\textrm{min}}\rho_s= 0.054$ and 12 $k_y$ grid points with $k_{y,\mathrm{min}}\rho_s=0.07$. The spectral wavenumber advection method was used to implement flow shear \cite{candy2018spectral}. Without any imposed $\mathrm{E\times B}$ shear, ion scale $k_y$ simulations of the MAST equilibrium saturate at a small level of flux, which is well below the experimental level\footnote{This equilibrium also contained electron scale ETG modes which contributed significantly more heat flux much closer to the experimental level, suggesting that the MTMs were not as experimentally relevant in the chosen radial position \cite{giacomin2023nonlinear}.}. The simulation shown in Figure \ref{fig:mast_nl_flux} has a predicted electron flux of $Q_e = (1.7\pm0.3) \times 10^{-2} Q_{gB}$, calculated by taking the mean value in the latter half of the simulation with the uncertainty given by the standard deviation. This dominates compared to the particle and ion heat flux which are $\Gamma_e = (2.4\pm0.4) \times 10^{-4} \Gamma_{gB}$ and $Q_i = (4.3\pm0.3) \times 10^{-4} Q_{gB}$ respectively. Furthermore, the transport is dominated by the electromagnetic contribution, with a time averaged $Q_e^{A_{||}} / Q_\mathrm{total} = 0.82$, indicating that it is indeed the MTM that is causing this transport, rather than the PEM This is further confirmed by running this case electrostatically, which results in transport two orders of magnitude smaller.
These simulations saturate with large zonal $\phi$ and $A_{||}$ which may be relevant to the saturation mechanism. Recent work by M. Giacomin \textit{et al} \cite{giacomin2023nonlinear}, which also examined this equilibrium, suggests that the stochastic transport, which is typically the dominant channel for MTMs, is weak in this region due to low magnetic shear resulting in a large separation between rational surfaces compared to the island width. A simulation was then performed adding in the experimental level of $\mathrm{E\times B}$ shear with $\gamma_{\mathrm{E\times B}}^\mathrm{exp} = 0.19 c_s/a$. The heat flux here is $Q_e=(2.8\pm0.3) \times 10^{-2} Q_{gB}$, which is similar to the case without $\mathrm{E\times B}$ shear. The impact of equilibrium $\mathrm{E\times B}$ shear on MTM turbulence is minimal, consistent with expectations from the weak dependence of $\gamma^\mathrm{MTM}$ on $\theta_0$, as discussed in Section \ref{sec:mast_lin}. %Note this is in stark contrast to the strongly stabilising impact of equilibrium $\mathrm{E\times B}$ shear on modes, like toroidal ITG and KBMs, that are more strongly ballooning in character.

\section{NSTX} \label{sec:nstx}

Here we assess whether $\beta_\mathrm{eff}$ is also a reliable indicator of linear MTM stability for the local equilibrium from NSTX with the local parameters in Table \ref{tab:miller_params}, taken from \cite{guttenfelder2012scaling}, where MTMs were also found. As with the MAST case, we assess the dependence of $\gamma^\mathrm{MTM}$ on $\theta_0$, and the impact of $\mathrm{E\times B}$ shear on the saturation level of the turbulence in nonlinear simulations.

\subsection{Linear simulations}

\begin{figure}[htb]
    \begin{subfigure}{0.49\textwidth}
        \centering
        \includegraphics[width=74mm]{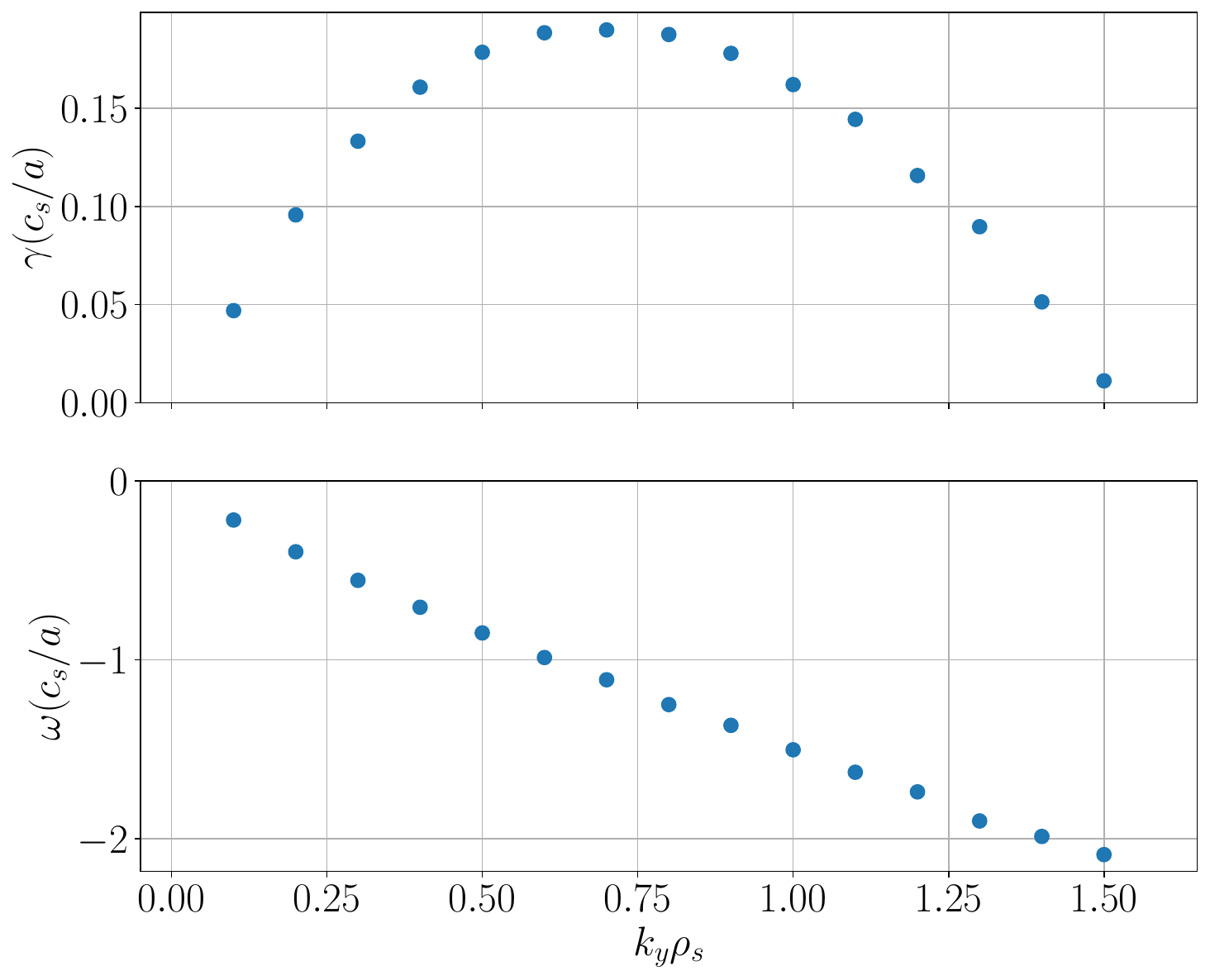}
        \caption{}
        \label{fig:nstx_eigval}
    \end{subfigure}
    \begin{subfigure}{0.49\textwidth}
        \centering
        \includegraphics[width=77mm]{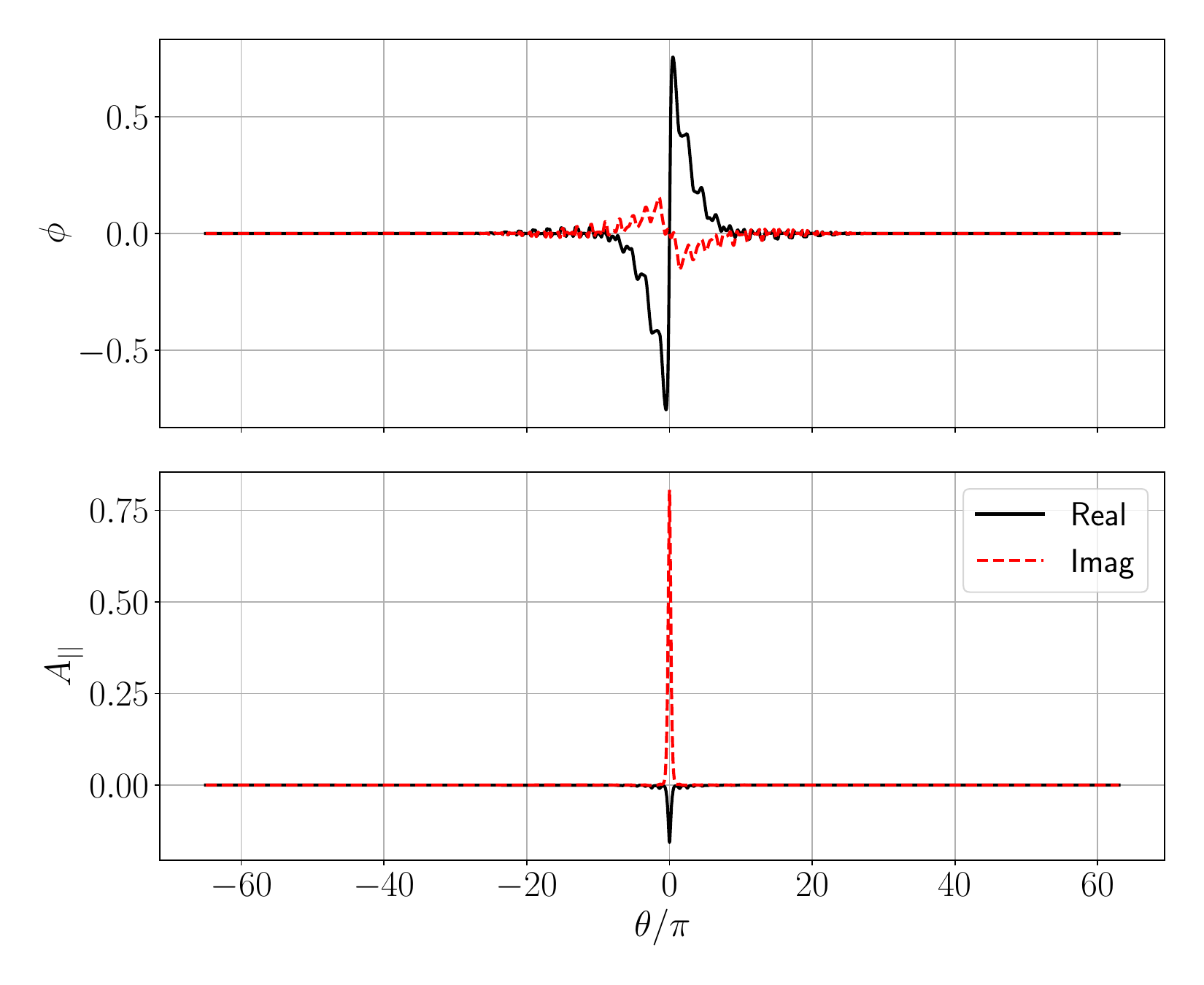}
        \caption{}
        \label{fig:nstx_eigfunc_ky05}
    \end{subfigure}
    \caption{a) Eigenvalues of the NSTX equilibrium at $\theta_0=0$. b) Eigenfunction of MTM at $k_y\rho_s=0.5$ in the NSTX equilibrium.}
    \label{fig:nstx}
\end{figure}

Figure \ref{fig:nstx_eigval} shows the growth rate and mode frequency for the MTMs found in NSTX. The results here match well with that shown in \cite{guttenfelder2012simulation}, with no electrostatic mode being seen here. Compared to MAST, the MTMs here have a much larger normalised 
growth rate so it is not surprising that they are unstable up to a higher $k_y\rho_s=1.5$. It might also be expected that any nonlinear simulation without sheared flow would saturate at a higher flux than for MAST. The eigenfunctions at $k_y\rho_s=0.5$ are shown in Figure \ref{fig:nstx_eigfunc_ky05}; the electrostatic potential is considerably less extended in ballooning space compared to Figure \ref{fig:mast_eigfunc_ky05}, due to both the higher collisionality and higher $\hat{s}$. 

\begin{figure}[!htb]
    \centering
    \includegraphics[width=80mm]{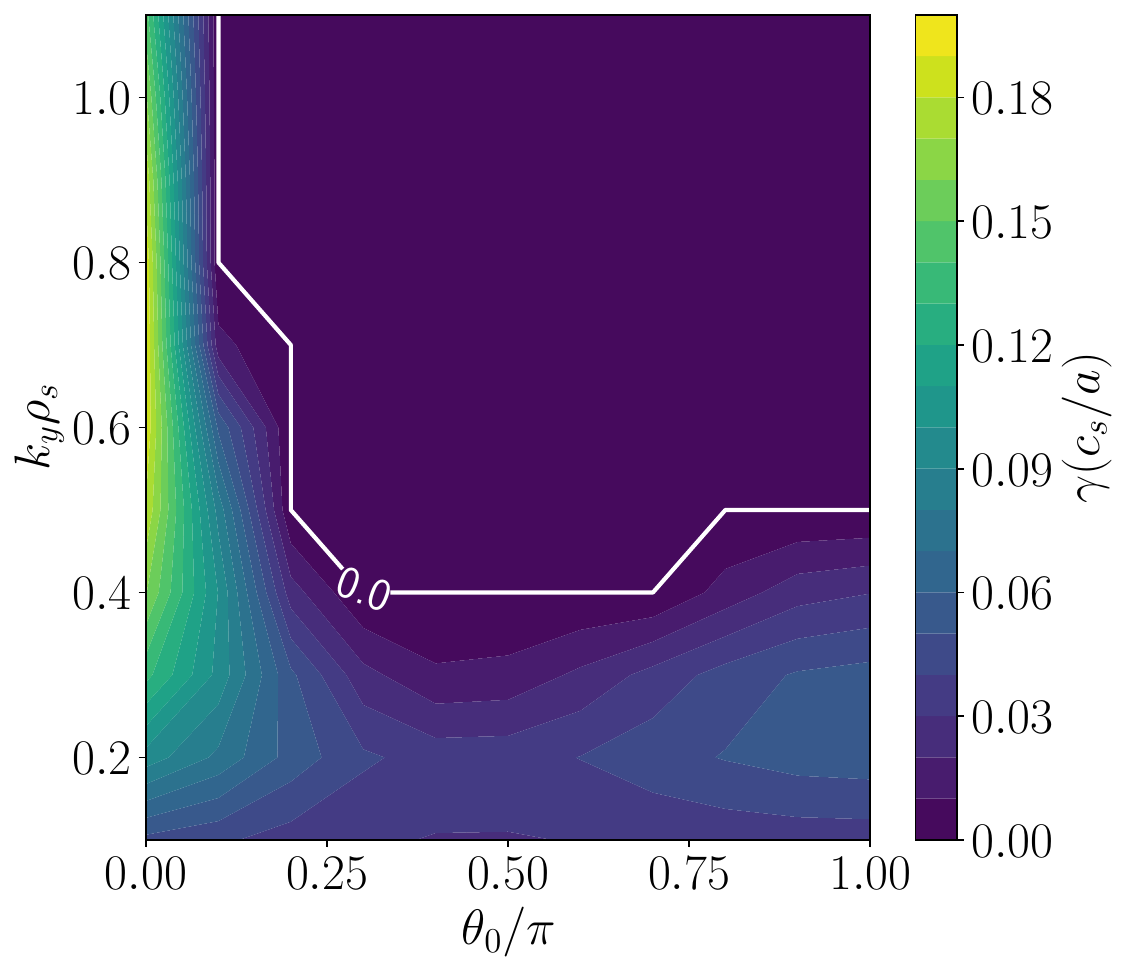}
    \caption{2D contour plot of the growth rate against $\theta_0$ and $k_y\rho_s$ for the NSTX local equilibrium. The solid white line denotes the marginal stability contour.}
    \label{fig:theta0_ky_scan_nstx_gam}
\end{figure}

A 2D linear stability scan has been performed in $k_y$ and $\theta_0$, similar to that performed in Section \ref{sec:mast}, spanning $k_y\rho_s$ from $0.1 \rightarrow 1.1$ and $\theta_0$ from $0 \rightarrow 2\pi$. Figure \ref{fig:theta0_ky_scan_nstx_gam} presents a contour plot of the MTM growth rate, where the white line shows the marginal stability contour. The only unstable mode found here was the MTM. For $k_y\rho_s \le 0.3$, the mode remains unstable for all values of $\theta_0$, but the growth rates are non-monotonic with $\theta_0$. At $k_y\rho_s=0.4$ a window of stability appears centred around $\theta_0=0.4\pi$, getting wider in $\theta_0$ at higher $k_y$, restricting the unstable space to a narrow region around $\theta_0=0.0$.

%This is expected to play minor role in the $\mathrm{E\times B}$ shear suppression of transport as at lower $k_y\rho_s$ the level of transport tends to be lower due to the increased distance between magnetic islands resulting in less stochasticity.

%\begin{figure}[!htb]
%    \centering
%    \includegraphics[width=80mm]{figures/NSTX_omega_theta0_cont.pdf}
%    \caption{.}
%    \label{fig:theta0_ky_scan_nstx_ome}
%\end{figure}

\begin{figure}[htb]
    \begin{subfigure}{0.49\textwidth}
        \centering
        \includegraphics[width=70mm]{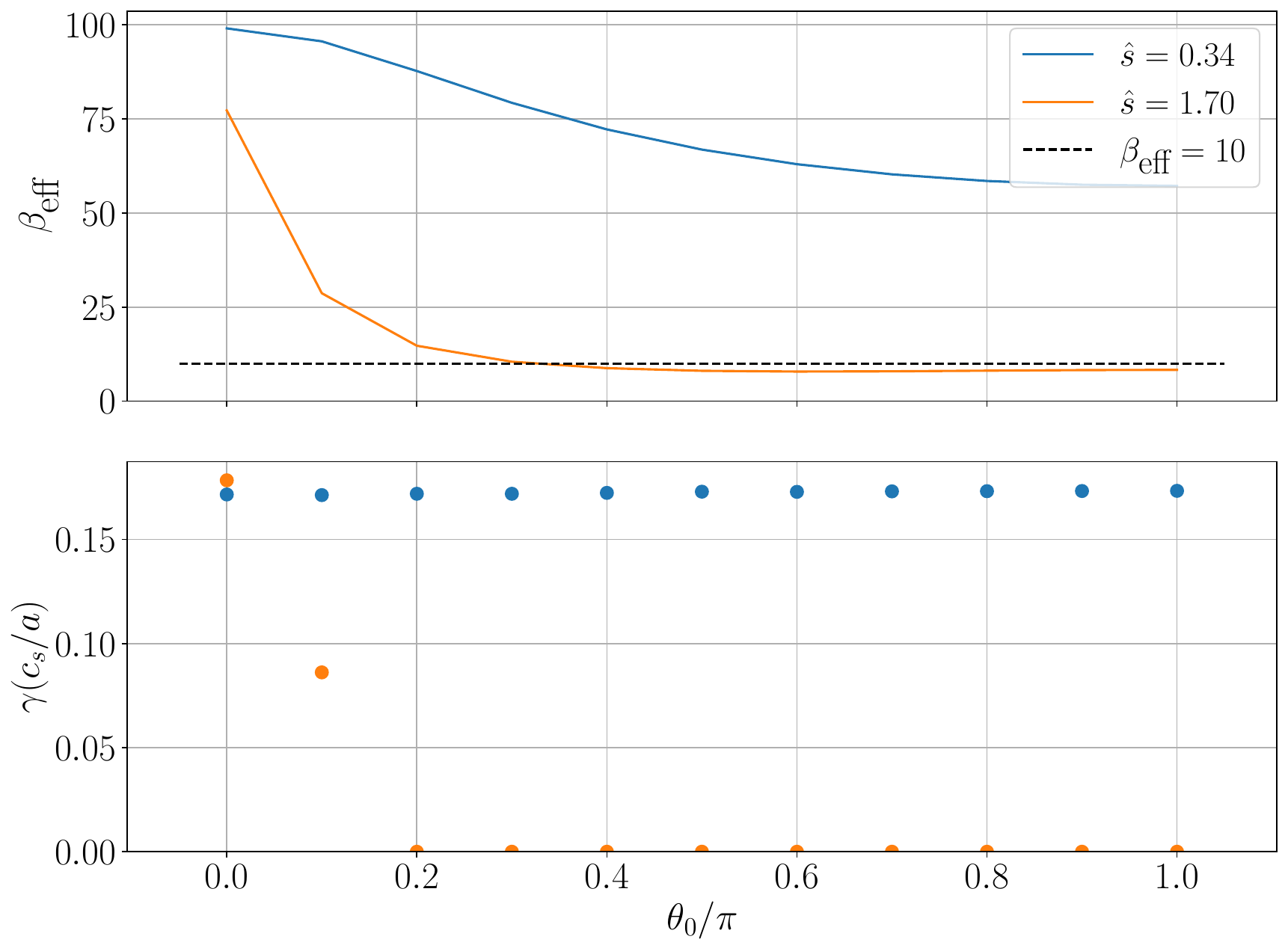}
        \caption{}
        \label{fig:gamma_beta_eff_nstx}
    \end{subfigure}
    \begin{subfigure}{0.49\textwidth}
        \centering
        \includegraphics[width=70mm]{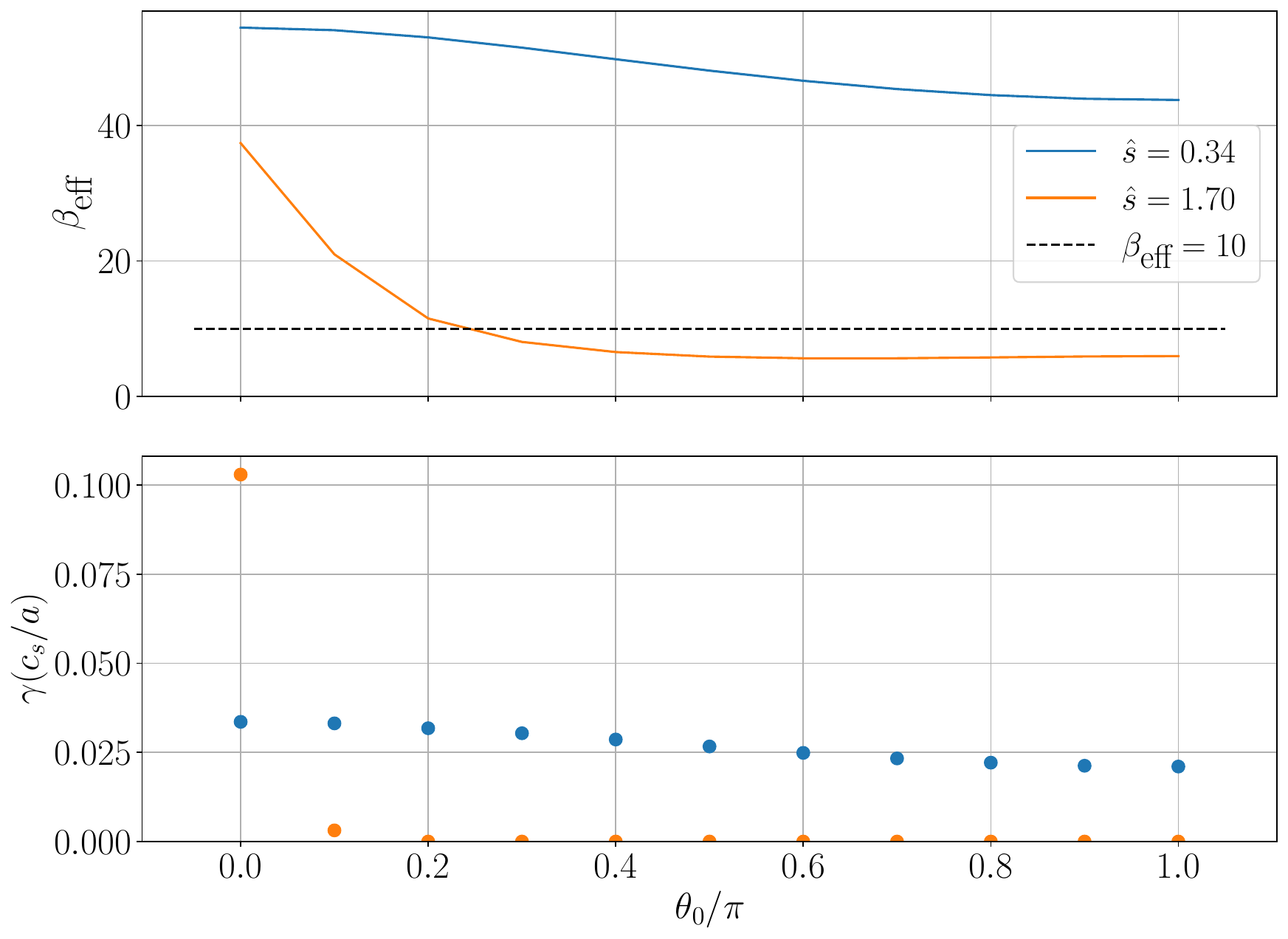}
        \caption{}
        \label{fig:gamma_beta_eff_mast}
    \end{subfigure}
    \caption{Comparing how $\beta_\mathrm{eff}$ (top) and $\gamma$ (bottom) change with $\theta_0$ for $k_y\rho_s=0.5$. (a) and (b) use the NSTX and MAST equilibria respectively. In both figures, simulations using the MAST $\hat{s}=0.34$ are shown in blue and the NSTX $\hat{s}=1.70$ are in orange. The black dashed line illustrates $\beta_\mathrm{eff}=10$.}
    \label{fig:gamma_beta_eff}
\end{figure}

The MAST and NSTX local equilibria show a very different dependence of $\gamma^\mathrm{MTM}$ on $\theta_0$, even though the values of many local parameters are quite similar. We have identified the local equilibrium parameters responsible for this striking difference by individually changing each equilibrium parameter from NSTX to that from MAST. This highlighted the magnetic shear, $\hat{s}$, as the most significant parameter. Figure \ref{fig:gamma_beta_eff_nstx} shows how the growth rate varies with $\theta_0$ for $k_y\rho_s=0.5$ using the NSTX equilibrium at two different values of $\hat{s}$. The orange line uses the NSTX equilibrium value of $\hat{s}=1.70$ and the MAST value of $\hat{s}=0.34$ is in blue. With the NSTX equilibrium $\hat{s}$, the mode is stable for $\theta_0>0.1\pi$, which coincides with $\beta_\textrm{eff}$ dropping below 10. At the lower MAST value of $\hat{s}$ the mode is unstable and $\beta_\textrm{eff}>50$ across the whole range in $\theta_0$.

\begin{figure}[!htb]
    \begin{subfigure}{0.49\textwidth}
        \centering
        \includegraphics[width=75mm]{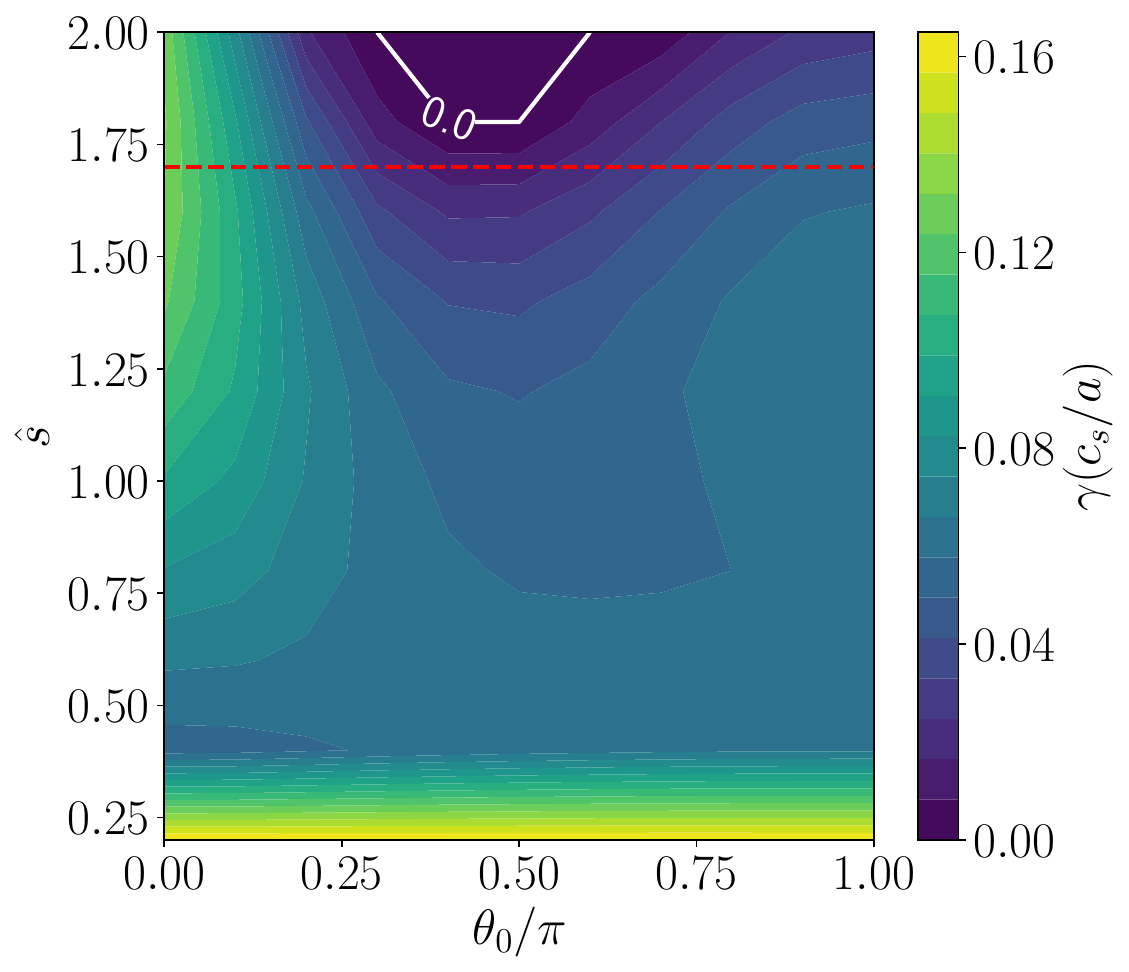}
        \caption{}
        \label{fig:gamma_shat_nstx_cont}
    \end{subfigure}
    \begin{subfigure}{0.49\textwidth}
        \centering
        \includegraphics[width=75mm]{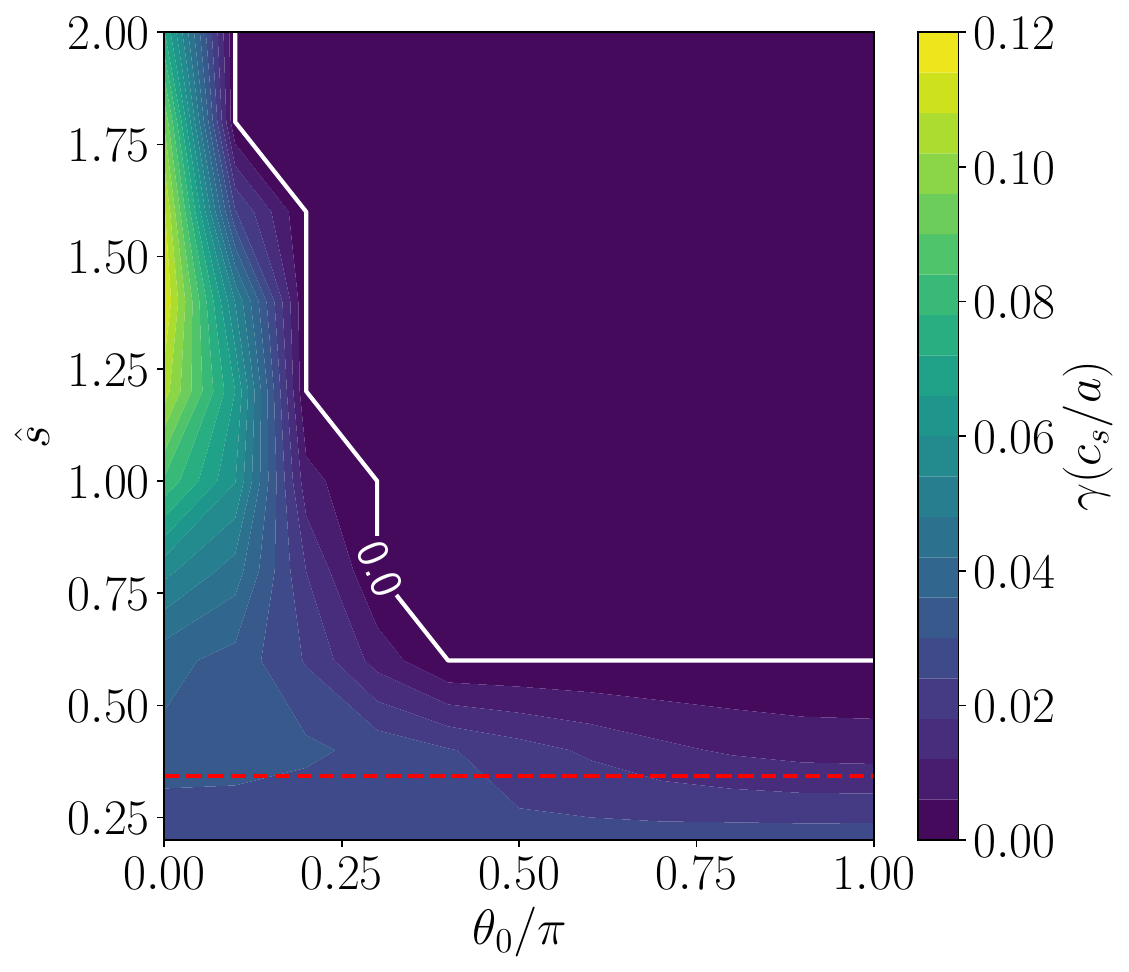}
        \caption{}
        \label{fig:gamma_shat_mast_cont}
    \end{subfigure}
    \caption{2D contour plot of the growth rate against $\theta_0$ and $\hat{s}$ for a) the NSTX equilibrium at $k_y\rho_s=0.3$ and b) the MAST equilibrium at $k_y\rho_s=0.5$. The dashed red line denotes the equilibrium $\hat{s}$ for that surface.}
    \label{fig:gamma_shat_cont}
\end{figure}

To further confirm the impact of $\hat{s}$, simulations were run for the MAST local equilibrium case in Section \ref{sec:mast} with the equilibrium MAST $\hat{s}$ in blue and the higher NSTX $\hat{s}$ in orange, with the results shown in Figure \ref{fig:gamma_beta_eff_mast}. At higher $\hat{s}$, $\gamma^\mathrm{MTM}$ becomes much more sensitive to $\theta_0$ and becomes stable at higher $\theta_0$. (Note the electrostatic PEM at $k_y\rho_s=0.5$ found in the MAST local equilibrium is stabilised with higher $\hat{s}$.) This suggests that MTMs found in regions with high magnetic shear may have their transport suppressed by $\mathrm{E\times B}$ shear. Figure \ref{fig:gamma_shat_cont} show 2D contour plots of $\gamma^\textrm{MTM}$ against $\theta_0$ and $\hat{s}$ for the MTM at $k_y\rho_s=0.3$ in NSTX and for the MTM at $k_y\rho_s=0.5$ in MAST, where it is clear that the dependence of $\gamma^\textrm{MTM}$ on $\theta_0$ is increasingly insensitive and monotonic at low values of $\hat{s}$, and that the unstable region with peak growth rate at $\theta_0=0$ narrows as $\hat{s}$ is increased. All the unstable modes found in these scans were MTMs.

As mentioned earlier, increasing $\hat{s}$ will make $\beta_\mathrm{eff}$ drop off faster with $\theta_0$, which according to the model will help to stabilise the mode. Figure \ref{fig:gamma_beta_eff} shows that $\beta_\textrm{eff}$ has a stronger dependence on $\theta_0$ at higher $\hat{s}$, dropping below $10$ by $\theta_0=0.3\pi$ for both equilibria. Note the higher $\hat{s}$ cases have a lower $\beta_\textrm{eff}$ but are more unstable at $\theta_0=\pi$. This is not inconsistent with the theory as $\beta_{e,\mathrm{crit}}$ will also change with $\hat{s}$. If this can be determined independently by a reduced model then it will be possible to determine at what $\theta_0$ the mode goes stable, which was shown in Figure \ref{fig:mast_beta_eff_scan}. The change in $\beta_\mathrm{eff}$ can be attributed to how $k_\perp$ increases along the field line in the two different equilibria as the larger NSTX $\hat{s}$ will result in $k_\perp$ becoming proportionally larger for a given ballooning angle.

%The change in $\beta_\mathrm{eff}$ can be attributed to how $k_\perp$ increases along the field line in the two different equilibria as the larger $\hat{s}$ in the NSTX case will result in $k_\perp$ becoming proportionally larger for a given ballooning angle.

\commentout{
\begin{figure}[!htb]
    \centering
    \includegraphics[width=80mm]{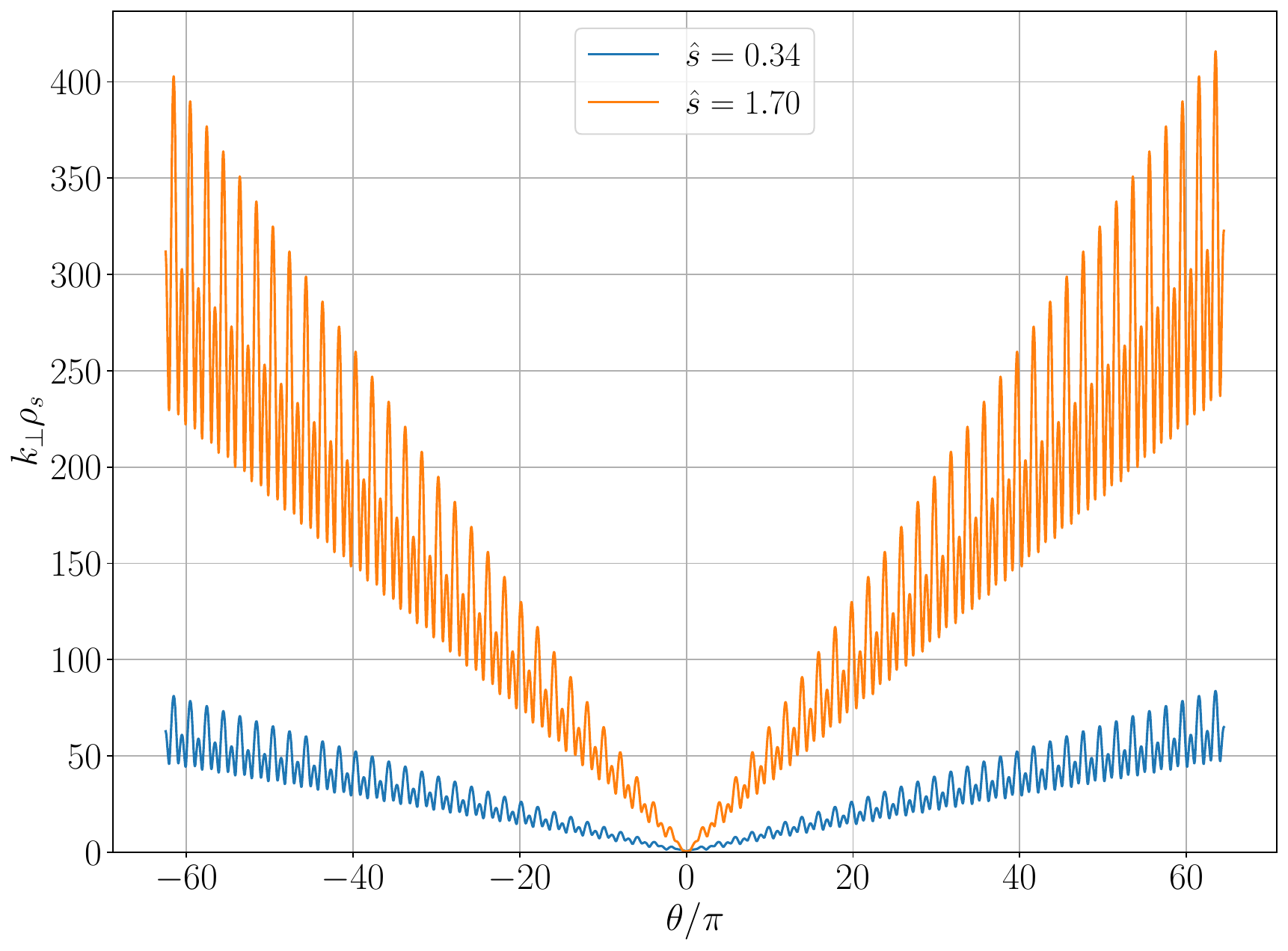}
    \caption{$k_\perp\rho_s$ is shown as a function of the ballooning angle for MAST with low $\hat{s}$ (blue) and high $\hat{s}$ (orange) for $\theta_0=0$.}
    \label{fig:mast_kperp}
\end{figure}

}
\begin{figure}[!htb]
    \centering
    \includegraphics[width=80mm]{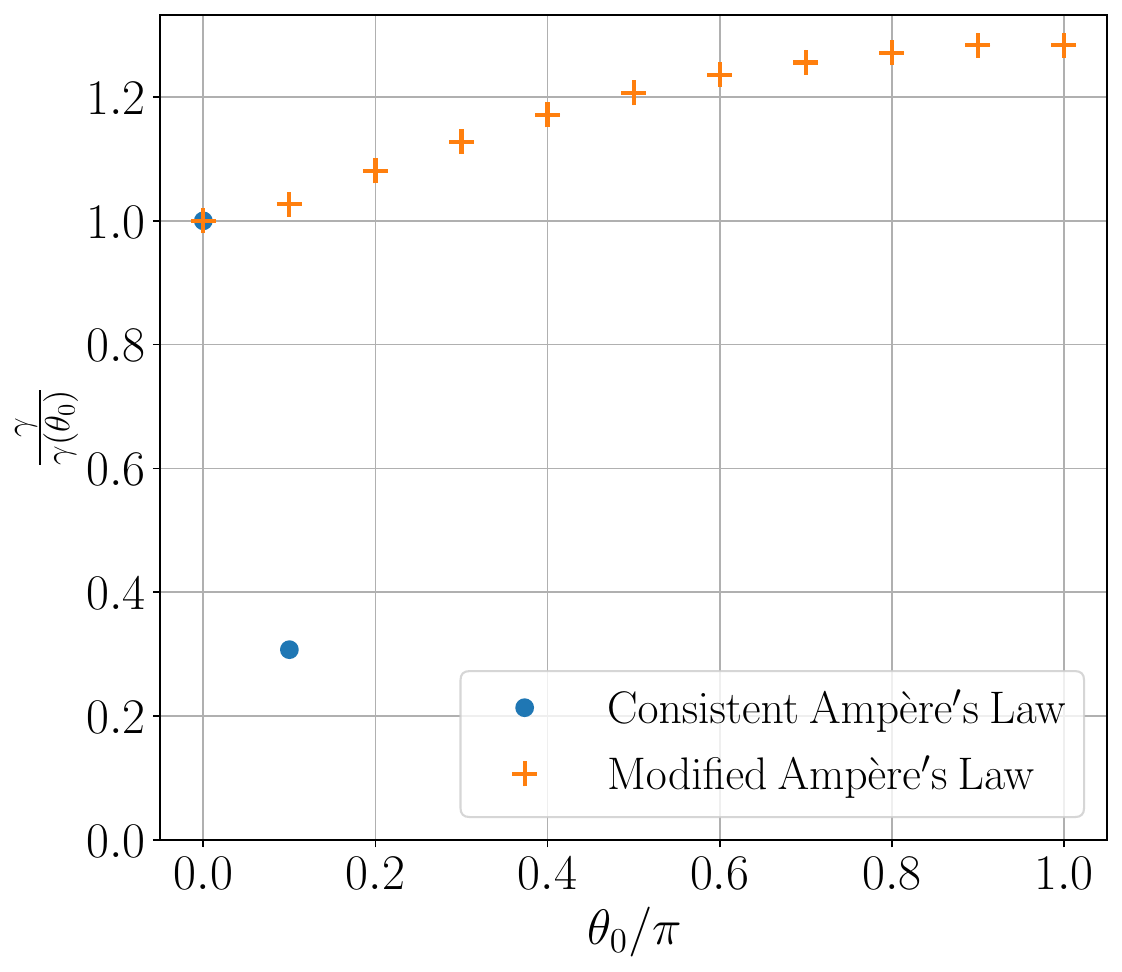}
    \caption{Growth rate of the MTM at $k_y\rho_s=0.6$ for NSTX equilibrium using the consistent $k_\perp\rho_s$ in Amp\`ere's law (blue) and the modified Amp\`ere's law using $k_\perp\rho_s$ with an artificially lower $\hat{s}$ (orange).}
    \label{fig:beta_ampere_change}
\end{figure}

\begin{figure}[!htb]
    \centering
    \includegraphics[width=80mm]{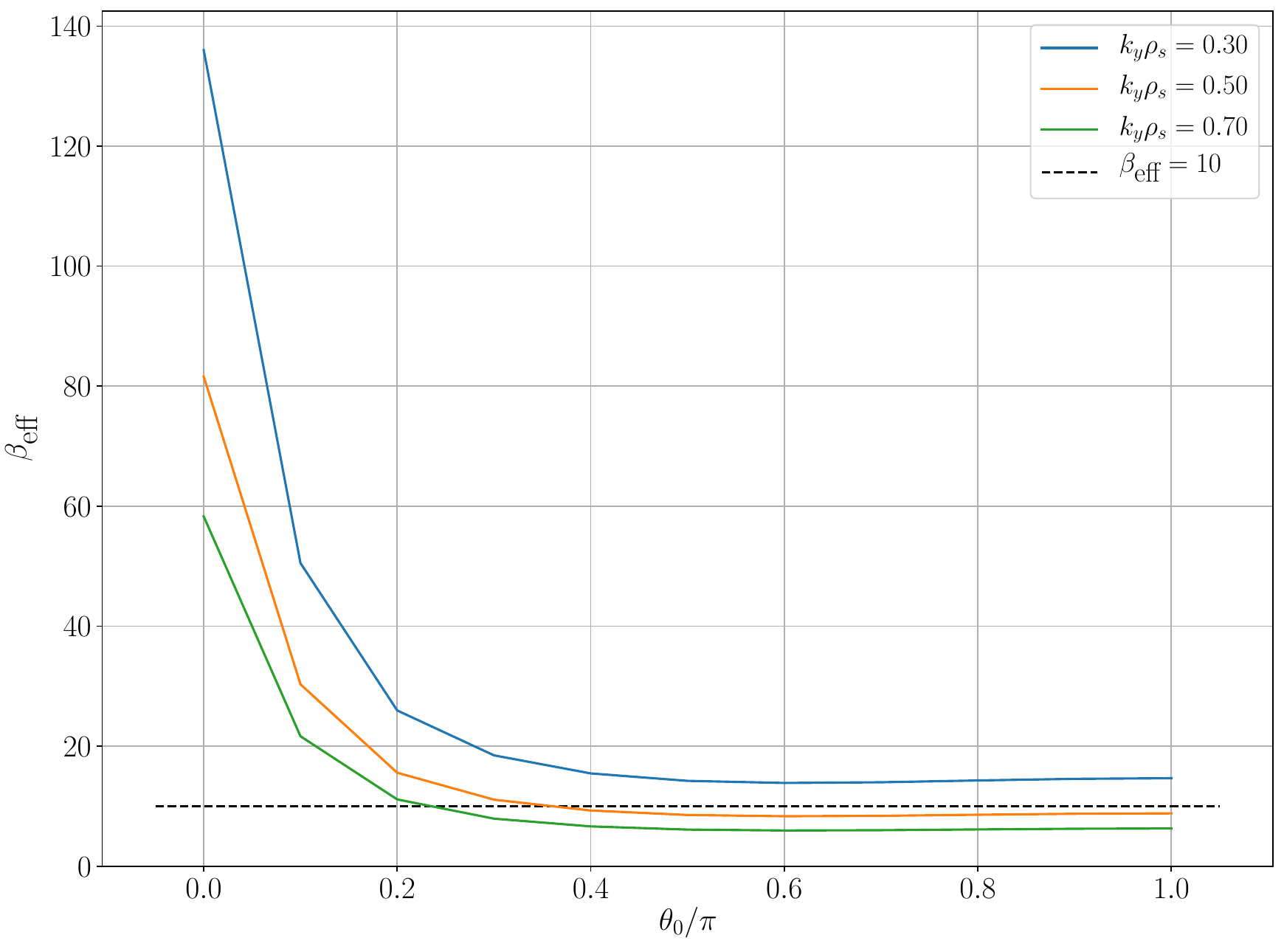}
    \caption{$\beta_{\mathrm{eff}}$ as a function of $\theta_0$ at different $k_y\rho_s$ for the NSTX equilibrium. The black dashed line illustrates $\beta_\mathrm{eff}=10$.}
    \label{fig:beta_eff_nstx}
\end{figure}

\begin{figure}[!htb]
    \begin{subfigure}{0.49\textwidth}
        \centering
        \includegraphics[width=75mm]{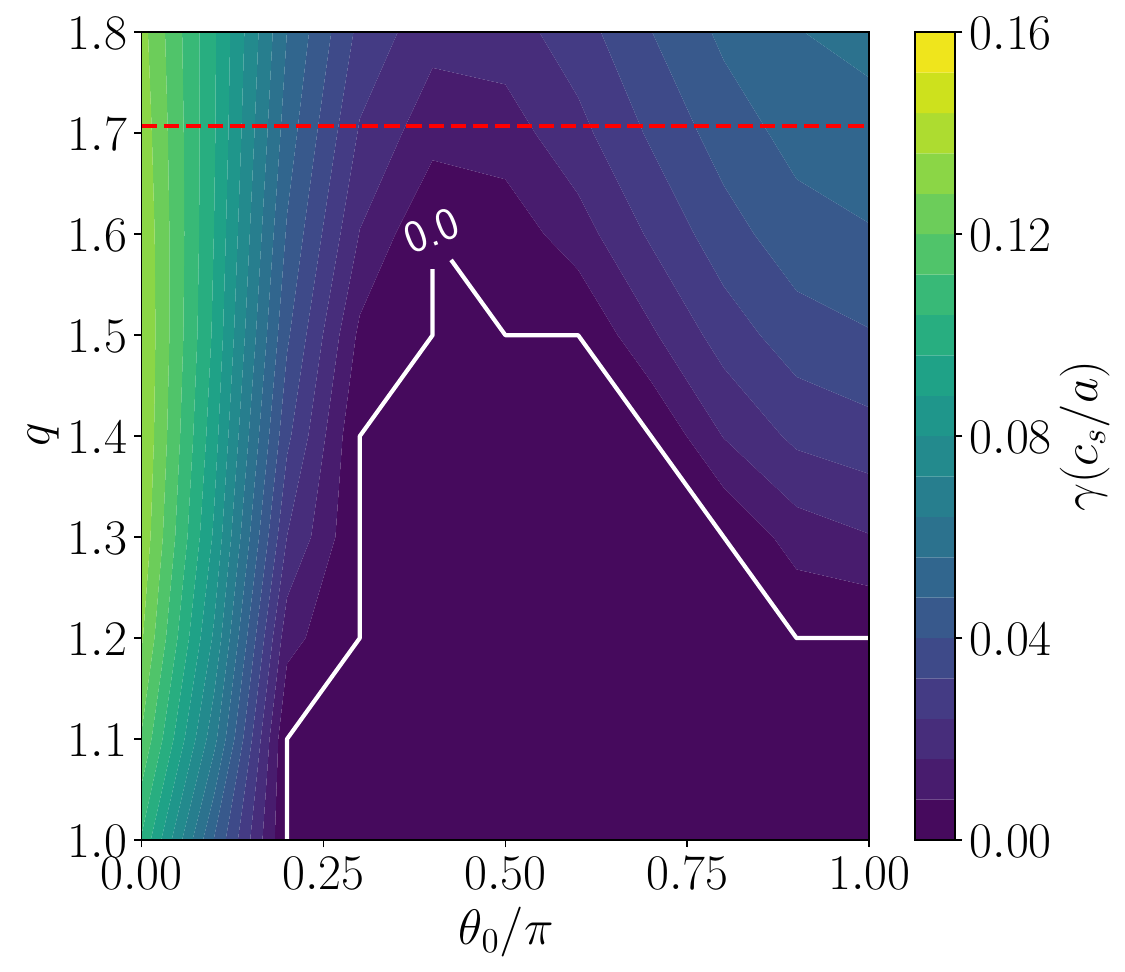}
        \caption{}
        \label{fig:gamma_q_nstx_cont}
    \end{subfigure}
   \begin{subfigure}{0.49\textwidth}
        \centering
        \includegraphics[width=75mm]{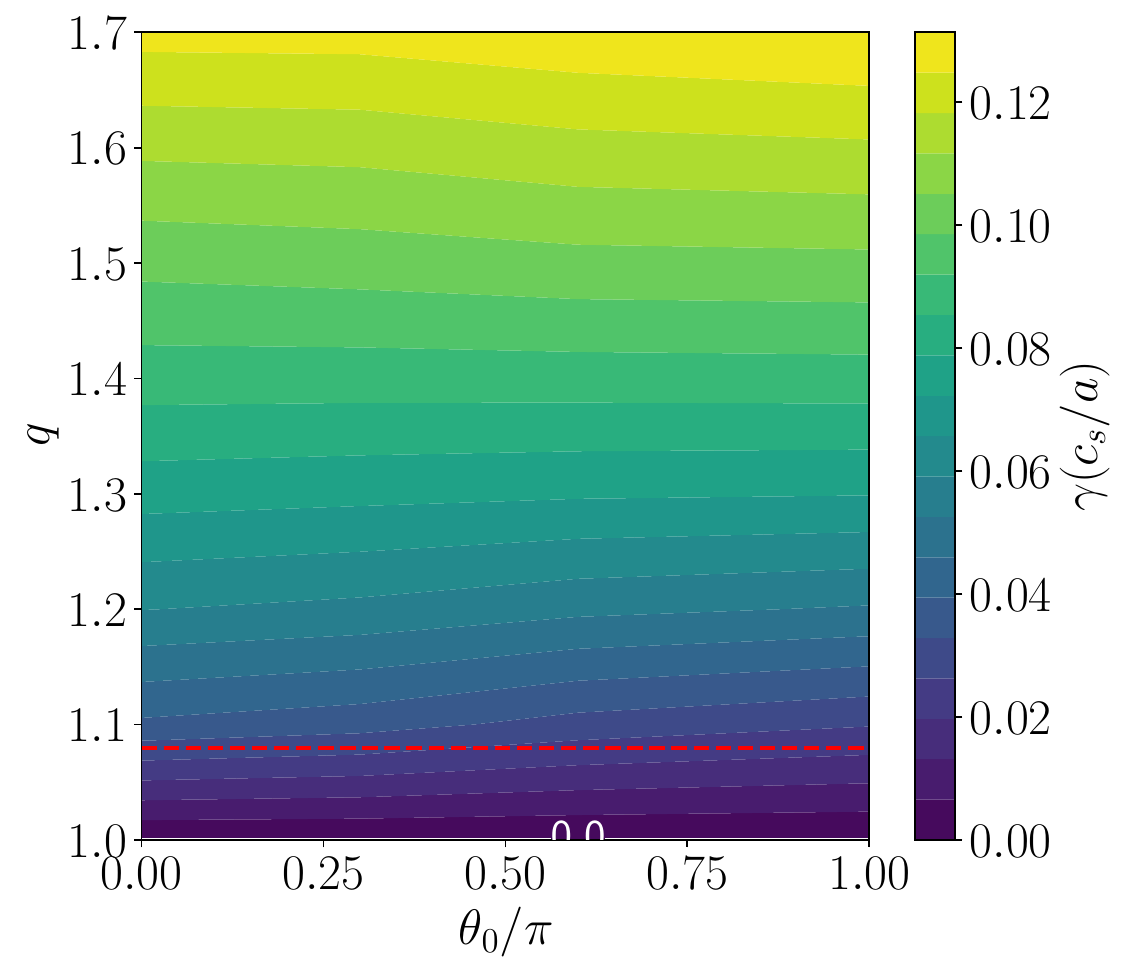}
        \caption{}
        \label{fig:gamma_q_mast_cont}
    \end{subfigure}
    \caption{2D contour plot of the growth rate against $\theta_0$ and $q$ for a) the NSTX equilibrium at $k_y\rho_s=0.3$ which has $\hat{s}=1.70$ and b) the MAST equilibrium at $k_y\rho_s=0.5$ which has $\hat{s}=0.34$. The dashed red line denotes the equilibrium $q$ for that simulation.}
    \label{fig:gamma_q_cont}
\end{figure}

Magnetic shear only appears within the definitions of $k_x$, $k_\perp$, the curvature drift and the grad-$B$ drift terms within gyrokinetic equation. To isolate which of one the impacts of changing $\hat{s}$ is responsible for the changes in the stability, several NSTX simulations were performed where $\hat{s}$ was artificially lowered to the MAST value independently in each place in the gyrokinetic equation where it appears. This revealed that the impact of $\hat{s}$ on $k_\perp$ is entirely the responsible for $\gamma^\mathrm{MTM}$ becoming insensitive to $\theta_0$ at low $\hat{s}$. In a more detailed refinement of this investigation focusing on the impact of magnetic shear on $k_\perp$, the change in the dependence of $\gamma^\mathrm{MTM}$ on $\theta_0$, illustrated in Figure \ref{fig:beta_ampere_change}, can be attributed directly to where $k_\perp$ enters Amp\`ere's law,\footnote{$\delta\mathbf{B_{\perp}} \propto 1/k_{\perp}$ from equation~\ref{eqn:amp}, so the perturbed field is increasingly localised in the parallel direction at higher $\hat{s}$ because $k_{\perp}$ increases more rapidly with $\theta$.} Modifying $k_\perp\rho_s$ in the NSTX local equilibrium to use the lower $\hat{s}$ value from MAST, the growth rate actually increases with $\theta_0$ (which is also found in high $q$ MAST simulations that will be shown later in Figure \ref{fig:gamma_q_mast_cont}). This confirms that it is specifically how high magnetic shear impacts Amp\`ere's law that allows $\theta_0$ stabilisation and thus for $\mathrm{E\times B}$ shear suppression to be effective. This provides further evidence that $\beta_\mathrm{eff}$ is a relevant parameter.

%\begin{figure}[!htb]
%    \centering
%    \includegraphics[width=80mm]{figures/NSTX_beta_eff_zoom.pdf}
%    \caption{.}
%    \label{fig:beta_eff_nstx_zoom}
%\end{figure}

The parameter $\beta_\mathrm{eff}$ is inversely proportional to $k_y$ which also helps explain why the MTM has a narrowing window of stability in $\theta_0$ (centred around $\theta_0=0.0$) at higher $k_y$. For the original NSTX case, $\beta_\textrm{eff}$ is shown for 3 different $k_y\rho_s$ in Figure \ref{fig:beta_eff_nstx}, and at higher $k_y\rho_s$, $\beta_\textrm{eff}$ is lowered. However, $k_y\rho_s=0.7$ has the lowest $\beta_\textrm{eff}$ at $\theta_0=0.0$, but is the most unstable out of the 3 $k_y\rho_s$ examined here. This indicates, unsurprisingly, that the linear growth rate is influenced by other parameters in addition to $\beta_\mathrm{eff}$, as are included in the parameter dependence derived in \cite{hardman2023new}.

However, this theory is not able to explain the behaviour at the lowest $k_y\rho_s$ where the growth rate is non-monotonic with $\theta_0$. For instance, in Figure \ref{fig:beta_eff_nstx} at $k_y\rho_s=0.3$, $\beta_\textrm{eff}$ is slightly non-monotonic with $\theta_0$, but not enough to explain the large growth rate found at $\theta_0=\pi$. This was different in the MAST equilibrium, where both $\gamma^\mathrm{MTM}$ and $\beta_\mathrm{eff}$ had a consistent monotonic dependence on $\theta_0$. In order to try to understand this, a scan in $q$ was performed for the NSTX local equilibrium. Figure \ref{fig:gamma_q_nstx_cont} shows a contour plot of $\gamma^\mathrm{MTM}(k_y \rho_s=0.3, \hat{s}=1.70)$ as a function of $q$ and $\theta_0$. In this scan MTMs are only unstable at $\theta_0=\pi$ when $q>1.2$. Whilst at lower $q$, $\gamma^\mathrm{MTM}$ decays monotonically with $\theta_0$, and there is no instability at $\theta_0=\pi$\footnote{All modes in Figure \ref{fig:gamma_q_nstx_cont} satisfy the MTM criterion $C_\mathrm{tear} > 0.1$.}.
Furthermore, Figure \ref{fig:gamma_q_mast_cont} shows a similar 2D scan for the MAST equilibrium where there is a relatively flat dependence of $\gamma^\mathrm{MTM}(k_y \rho_s=0.5, \hat{s}=0.34)$ on $\theta_0$ at low $q$, which becomes slightly peaked at $\theta_0=\pi$ at higher $q$. 

The peaking of $\gamma^\mathrm{MTM}$ at $\theta_0=\pi$, found in the above gyrokinetics simulations at higher $q$, is not captured in Hardman's model \cite{hardman2023new}. This is likely due to its low $\beta$ ordering assumptions, and in particular its neglect of the perturbed perpendicular current $J_\perp$, breaking down at higher $q$. In the model, $J_\perp$ is excluded in the charge continuity equation, as shown in Equation \ref{eqn:j_par}. However, the ratio of $(\nabla \cdot J_\perp) / (\nabla \cdot J_{||}) \propto \beta \big(\frac{qR}{a}\big)^2$, so increasing these terms makes $J_\perp$ term larger which violates this ordering. This indicates that changes to $\beta$ or $(R/a)^2$ should have a similar impact to changes in $q^2$. 

\begin{figure}[!htb]
    \begin{subfigure}{0.49\textwidth}
        \centering
        \includegraphics[width=75mm]{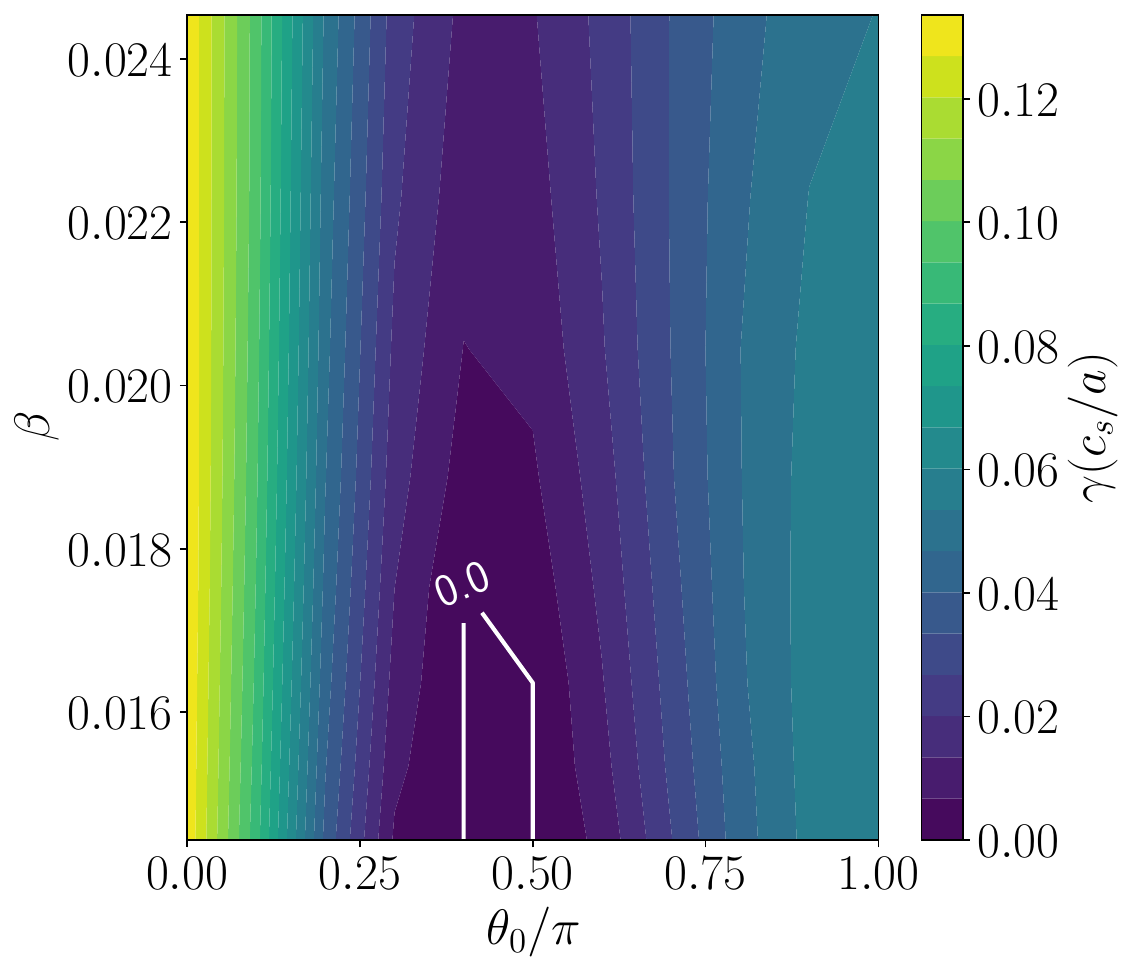}
        \caption{}
        \label{fig:gamma_beta_nstx_qR}
    \end{subfigure} 
    \begin{subfigure}{0.49\textwidth}
        \centering
        \includegraphics[width=75mm]{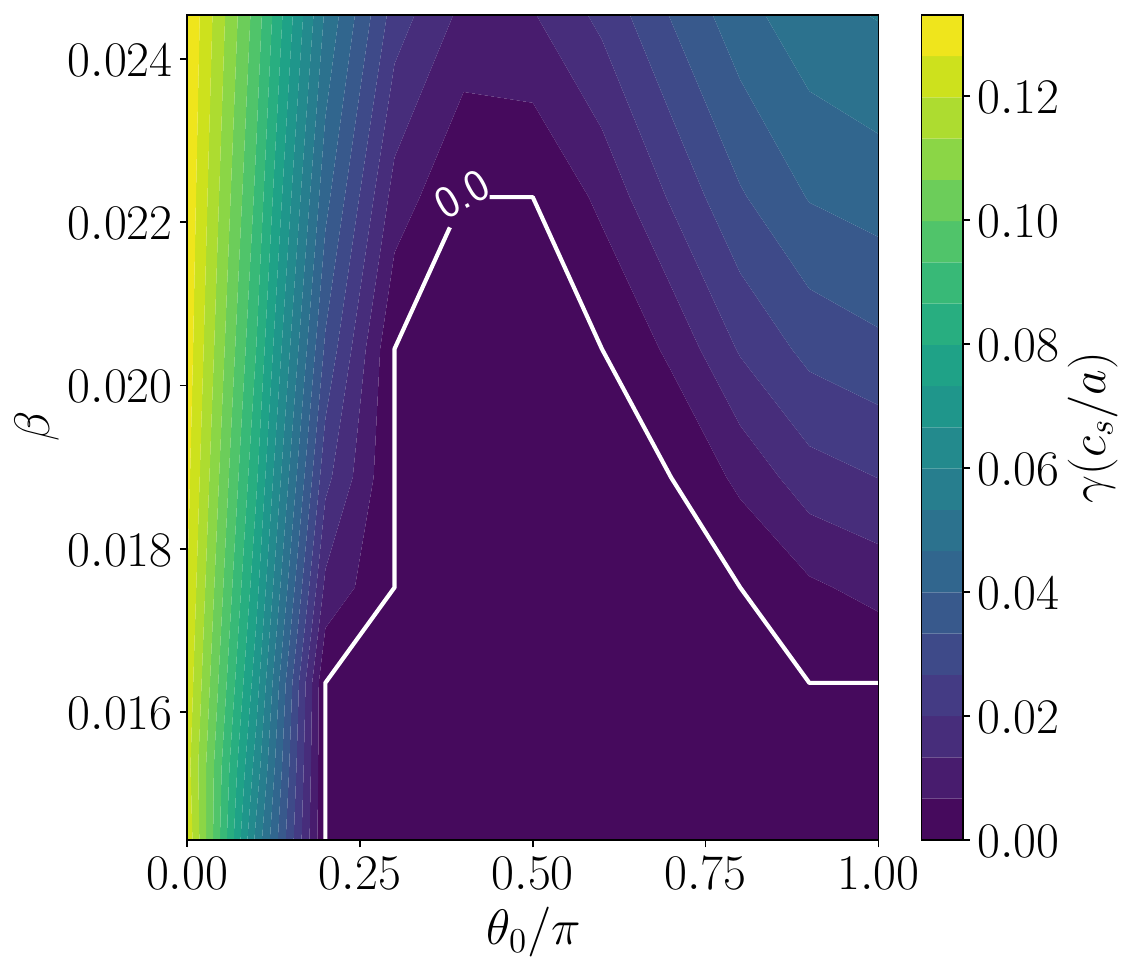}
        \caption{}
        \label{fig:gamma_beta_nstx}
    \end{subfigure}
    \caption{2D contour plot of the growth rate for the NSTX equilibrium against $\theta_0$ when a) changing $\beta$ whilst fixing $\beta (qR/a)^2$ by adjusting $R/a$ and b) changing $\beta$ at fixed $(qR/a)^2$. }
    \label{fig:gamma_beta_cont_qR}
\end{figure}

\begin{figure}[!htb]
    \begin{subfigure}{0.49\textwidth}
        \centering
        \includegraphics[width=75mm]{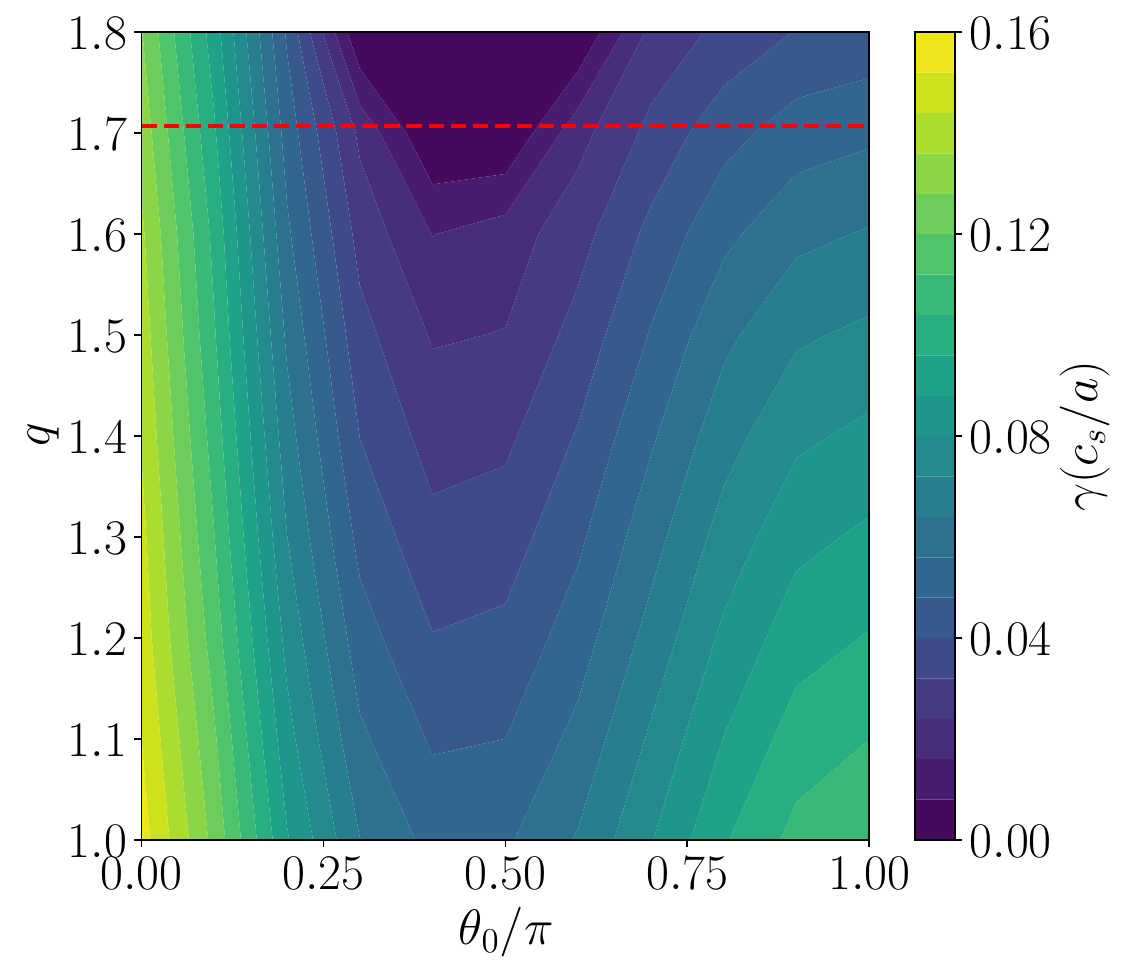}
        \caption{}
        \label{fig:gamma_q_nstx_qR}
    \end{subfigure} 
    \begin{subfigure}{0.49\textwidth}
        \centering
        \includegraphics[width=75mm]{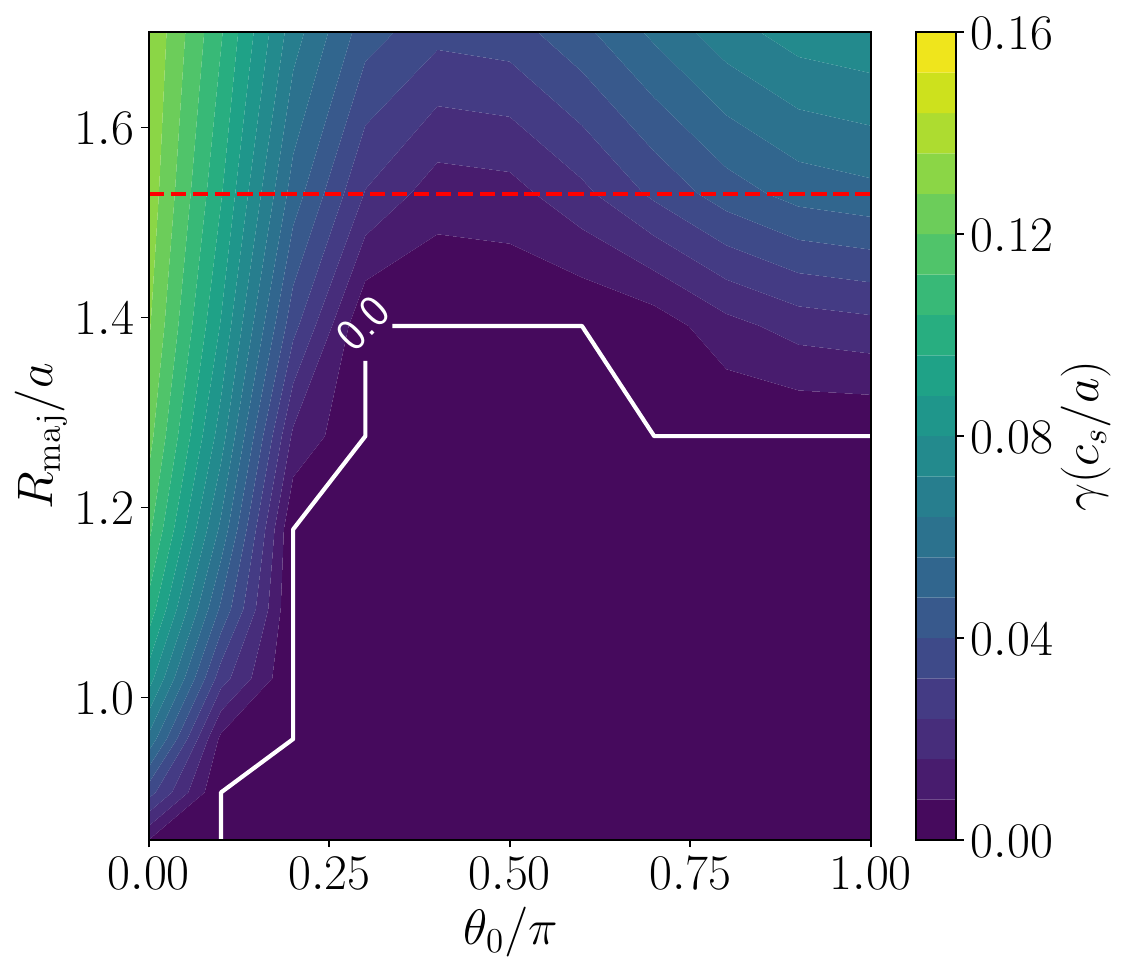}
        \caption{}
        \label{fig:gamma_R_nstx}
    \end{subfigure}
    \caption{2D contour plot of the growth rate for the NSTX equilibrium against $\theta_0$ when a) changing $q$ whilst fixing $\beta (qR/a)^2$ by adjusting $R/a$ and b) changing $R/a$ at fixed $\beta q^2$. The dashed red line denotes the equilibrium value for that simulation.}
    \label{fig:gamma_q_cont_qR}
\end{figure}

To test this, two additional sets of scans were performed changing $\beta$ and $R/a$ shown in Figures \ref{fig:gamma_beta_cont_qR} and \ref{fig:gamma_q_cont_qR} respectively. For both parameters, scans were performed in two ways. Firstly whilst maintaining a fixed $\beta \big(\frac{qR}{a}\big)^2$, would maintain the relative size of $J_\perp$. Here we see in both Figures \ref{fig:gamma_beta_nstx_qR} and \ref{fig:gamma_q_nstx_qR} that $\gamma^\textrm{MTM}$ remains non-monotonic throughout, contradicting the model as expected. Secondly, scans were performed whilst allowing $\beta \big(\frac{qR}{a}\big)^2$ to change, shown in Figures \ref{fig:gamma_beta_nstx} and \ref{fig:gamma_R_nstx}, which modifies the size of $J_\perp$ in a similar manner to the previous $q$ scan. Here the model's prediction is recovered as $\beta \big(\frac{qR}{a}\big)^2$ is reduced, similar to the lowering of $q$, which provides further evidence that the NSTX equilibrium is pushing beyond the orderings of the model.

%Two additional scans were performed, firstly in $R/a$ which modifies increases the size of $J_\perp$ in a similar way to $q$, illustrated in Figure \ref{fig:gamma_R_nstx}. Similar to $q$, as $R/a$ is reduced the monotonic behaviour of $\gamma^\mathrm{MTM}$ with $\theta_0$ is recovered. Figure \ref{fig:gamma_q_nstx_qR} shows the second scan in $q$ was performed whilst maintaining a constant $qR/a$. which would maintain the relative size of $J_\perp$. Here $\gamma^\textrm{MTM}$ remains non-monotonic throughout, which providing further evidence that the NSTX equilibrium is pushing beyond the orderings of the model.

%{\color{red} CMR think this para is out of order and already almost said... delete??? \sout{Even so, it appears that a larger value of $\hat{s}$ results in a larger variation of $\gamma^\mathrm{MTM}$ with $\theta_0$, potentially allowing for a region of stability to appear, even at higher $q$. This suggests that an equilibrium could still be optimised to allow for $\mathrm{E\times B}$ shear suppression.}}

\commentout{A potential reason for this is that as $q$ is increased, the relative size of the second peak to the main peak in $A_{||}$ at $\theta=0$ increases, displayed in Figure \ref{fig:gamma_q_mast_eigfunc}. The model assumes that $A_{||}$ is largest around $\theta=0$, but this is violated at higher $q$. It may be necessary to add an additional boundary layer in the model to account for this effect.
}

%This theory is valid up to $\beta_{\textrm{eff}} << (m_i/m_e)^\frac{1}{2} \sim 60$. Plotting $\beta_\textrm{eff}$ for the NSTX equilibrium we can see that for $\beta_{\textrm{eff}}$ drops off very quickly for this equilibrium, suggesting the drive for the MTM does decrease as $\theta_0$ increases. Furthermore, it is clear that as $k_y\rho_s$ increases, the total $\beta_{\textrm{eff}}$ decreases. At the highest $k_y\rho_s$, the mode becomes stable which is in line this theory. When $k_y\rho_s=0.5$, $\beta_\textrm{eff}$ approaches 10 and where the MTM remains unstable though weakly. However, the initial drop off in $\theta_0$ is in line with the change in $\beta_\textrm{eff}$ suggesting that even if we are a small amount over the $\beta_\textrm{eff}$, the theory may still partially hold. There is a limit to this as at the lowest $k_y\rho_s$, the $\beta_\textrm{eff}$ is much higher than the theory allows for all $\theta_0$ suggesting that we are beyond the region of validity. 

%The non-monotonic behaviour arises from the minimum $k_\perp$ being non-monotonic with $\theta_0$. Figure \ref{fig:nstx_gam_kperp} shows how this minimum lines up well with the minimum in the growth rate. This effect will be exacerbated with higher magnetic shear and shaping. If $k_{\perp,\textrm{min}}$ was monotonic, then the it should be expected that the growth rate would be monotonic.

\subsection{Nonlinear simulations}

\begin{figure}[!htb]
    \centering
    \includegraphics[width=80mm]{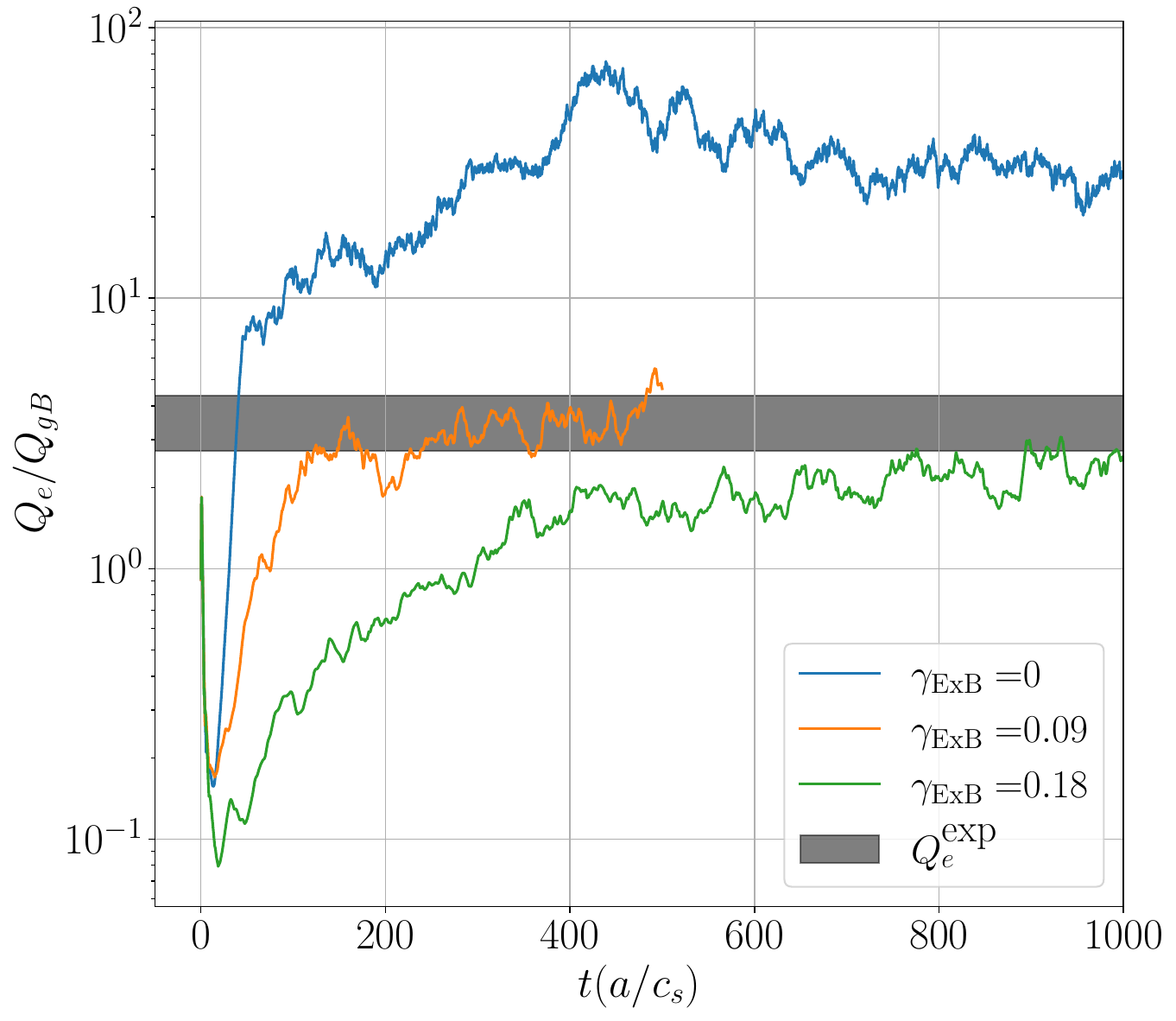}
    \caption{Nonlinear electron heat flux for NSTX simulations with varying levels of $\mathrm{E\times B}$ shear. Here $\gamma_\mathrm{E\times B}^{\mathrm{exp}} = 0.18 c_s/a$. Note that the electron heat flux dominated the total flux driving $>96\%$ of the total heat transport in these simulations. The grey band denotes the experimentally measured anomalous heat flux. The average and uncertainty in the flux is calculated from the final 50\% of time from each simulation.}
    \label{fig:exb_scan_nstx}
\end{figure}

Figure \ref{fig:theta0_ky_scan_nstx_gam} shows that a large region of the $(k_y\rho_s, \theta_0)$ phase-space is stable in the reference equilibrium of NSTX, suggesting that $\mathrm{E\times B}$ shear may help suppress the MTM transport. Nonlinear simulations used 256 $k_x$ grid points with a $k_{x,\textrm{min}}\rho_s= 0.068$ and 12 $k_y$ grid points with $k_{y,\mathrm{min}}\rho_s=0.07$ to perform a scan in $\gamma_\mathrm{E\times B}$, with the experimental value $\gamma_\mathrm{E\times B}^\textrm{exp}=0.18 c_s/a$. 
Figure \ref{fig:exb_scan_nstx} shows the level of electron heat flux for 3 different nonlinear simulations. When $\gamma_\mathrm{E\times B} = 0.0$, the simulation was found to saturate around $Q_e=(34\pm 7) Q_{gB}$\footnote{ Note that the same simulation was previously found to not saturate in \cite{patel2021confinement}. Here numerical instabilities that were responsible have been avoided through a recent improvement to the CGYRO parallel dissipation scheme - git commit \href{https://github.com/gafusion/gacode/commit/903307e3a56e306cdd6d211f92223424b7b98c16}{903307e}}; This is significantly higher than the MAST case and can be attributed to the higher MTM growth rates, together with a higher $q$ and $\hat{s}$, reducing the separation between adjacent rational surfaces and enhancing electron heat transport from stochastic magnetic fields \cite{giacomin2023nonlinear, rechester2020electron}.
At $\gamma_\mathrm{E\times B} =0.5 \gamma^\mathrm{exp}_\mathrm{E\times B}= 0.09 c_s/a$, the simulation was found to saturate at $Q_e=(3.5\pm0.5) Q_{gB}$, which is within the error of the experimental turbulent heat flux, shown in the shaded grey area. With the full $\gamma_\mathrm{E\times B} =\gamma^\mathrm{exp}_\mathrm{E\times B}= 0.18 c_s/a$, fluxes drop even further to $Q_e=(2.1\pm0.4) Q_{gB}$, slightly below the experimental value, though this likely lies within the uncertainty of $\gamma^\mathrm{exp}_\mathrm{E\times B}$ \footnote{We note that these CGYRO simulations are arguably more consistent with NSTX data than previously published MTM simulations using GYRO, where including $\mathrm{E\times B}$ shear resulted in $Q_e \ll Q_e^\mathrm{exp}$ \cite{guttenfelder2011electromagnetic} and the experimental heat flux could only be matched if $\mathrm{E\times B}$ was neglected.}.
The nonlinear simulations of Figure \ref{fig:exb_scan_nstx} demonstrate that when $\gamma^\mathrm{MTM}$ has a strong dependence on $\theta_0$, $\mathrm{E\times B}$ shear can be effective in suppressing MTM transport; in this NSTX case the electron heat flux reduces by more than an order of magnitude. 

\begin{figure}[!htb]
    \centering
    \includegraphics[width=80mm]{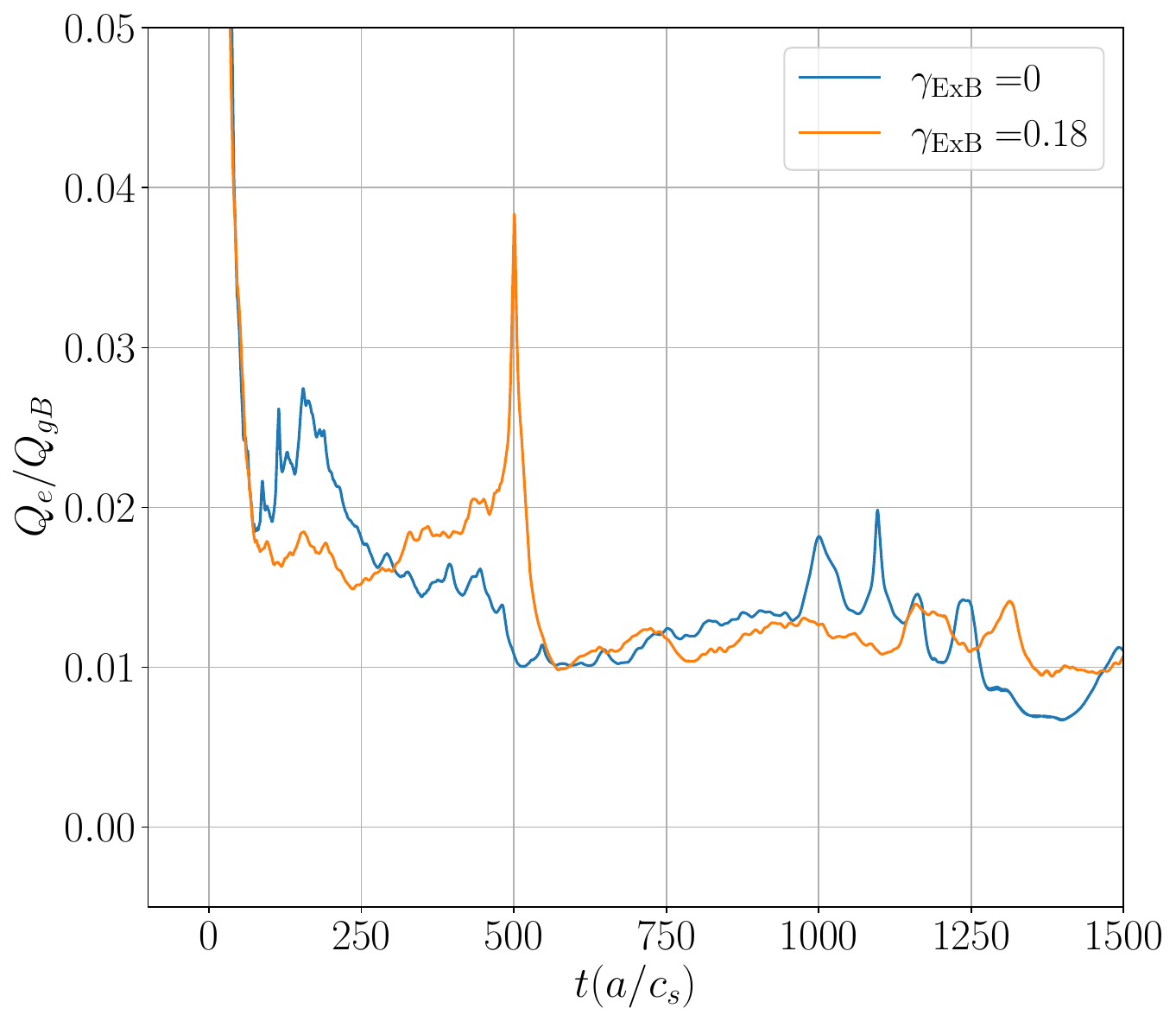}
    \caption{Nonlinear electron heat flux for NSTX simulations with $\hat{s}=0.34$ with varying levels of $\mathrm{E\times B}$ shear. Here $\gamma_\mathrm{E\times B}^{\mathrm{exp}} = 0.18 c_s/a$. Note that the electron heat flux dominated the total flux driving $>94\%$ of the total heat transport in these simulations.}
    \label{fig:exb_scan_nstx_lows}
\end{figure}

Note that the suppression of MTM turbulence was not observed in the nonlinear simulations of Figure \ref{fig:mast_nl_flux}, for the MAST surface at lower $\hat{s}$ where $\gamma^\mathrm{MTM}$ is insensitive to $\theta_0$. Even without flow shear, however, the absolute fluxes are extremely modest on this MAST surface due to the increased distance between rational surfaces at lower $\hat{s}$ \cite{giacomin2023nonlinear}.  Figure \ref{fig:exb_scan_nstx_lows} shows a nonlinear simulation for the NSTX surface, but using the lower value of $\hat{s}=0.34$ taken from the MAST surface: it is clear that the impact of $\mathrm{E\times B}$ shear is also minimal here. 

Thus we can conclude that $\mathrm{E\times B}$ shear suppression of MTM turbulence is more effective when $\gamma^\mathrm{MTM}$ is more strongly ballooning, which is favoured at higher $\hat{s}$.

%Note these simulations had $Z_\textrm{eff}=1$, compared to $2.9$ in the experiment which has been shown to destabilise MTM and may increase the flux further \cite{patel2021confinement, guttenfelder2012scaling}.

\section{Conclusion}
This work has helped understand the local plasma equilibrium conditions under which MTM transport should be more susceptible to suppression by perpendicular $\mathrm{E\times B}$ sheared flows, and shown that this can be identified linearly from the $\theta_0$ dependence of the linear growth rate. 

A recent linear theory of MTMs, Hardman \textit{et al} \cite{hardman2023new}, valid for $\beta_e \sim \sqrt{m_e/m_i}$, shows that the MTM growth rate depends on the parameter $\beta_\mathrm{eff}(\theta_0)$, and in this paper we have used local gyrokinetic calculations to compare $\gamma^\mathrm{MTM}(\theta_0)$ with $\beta_\mathrm{eff}(\theta_0)$ for a number of local equilibria from ST plasmas. 

In a local MAST equilibrium with $q=1.08$ and $\hat{s}=0.34$, $\gamma^\mathrm{MTM}$ is weakly dependant on $\theta_0$ and $\beta_\mathrm{eff}(\theta_0)$ follows a similar trend. Parameter scans demonstrate that $\gamma^\mathrm{MTM}$ is a unique function of $\beta_\mathrm{eff}$, as predicted by Hardman \textit{et al}, indicating that the theory captures the key properties of these linear modes. Nonlinear simulations of this equilibrium confirmed that $\mathrm{E\times B}$ shear had little impact on the predicted transport, in line with the weak dependence of $\gamma^\mathrm{MTM}$ on $\theta_0$.

In an NSTX local equilibrium with higher safety factor, $q=1.71$, and higher magnetic shear, $\hat{s}=1.70$, the MTMs have larger growth rates and are unstable up to a higher $k_y$. For $k_y \rho_s>0.5$, $\gamma^\mathrm{MTM}$ is unstable over a narrow window around $\theta_0=0$, and the growth rate drops steeply as $\theta_0$ increases, with $\beta_\mathrm{eff}(\theta_0)$ having a very similar character. A more detailed study demonstrates that this is due to $k_\perp$ increasing more rapidly along the field line at higher $\hat{s}$ (or finite $\theta_0$); this limits the parallel extent of $A_{||}$ (from Amp\`ere's law) and the radial displacement of the perturbed magnetic field that provides the linear drive. 

At lower $k_y$, however, MTMs become unstable with an additional peak in $\gamma^\mathrm{MTM}(\theta_0)$ at $\theta_0=\pi$ and this feature is not captured by the theory. In this theory the contributions from $J_\perp$ are excluded using the low $\beta$ ordering. However, this term is related to the size of $\beta \big(\frac{qR}{a}\big)^2$ and therefore a low $\beta$  can be offset by a higher $\frac{qR}{a}$. Scans were shown illustrating that as $\beta \big(\frac{qR}{a}\big)^2$ and thus the relative size of $J_\perp$ is reduced, the second peak on the inboard side disappears, indicating a breaking of the orderings that can occur at higher $q$ or $R/a$, even at low $\beta$, like in the NSTX equilibrium. 

\setcounter{footnote}{0} 

In nonlinear simulations, $Q_e$ matches the experimental flux when equilibrium $\mathrm{E\times B}$ shear is included\footnote{This improves on previous simulations that only matched experiment if $\mathrm{E\times B}$ shear was neglected \cite{guttenfelder2012simulation}.}, and $Q_e$ is an order of magnitude lower than the result of the $\gamma_\mathrm{E\times B}=0$ simulation. This mitigation by $\mathrm{E\times B}$ shear of the nonlinear MTM heat flux, is as expected given the strong dependence of $\gamma^\mathrm{MTM}$ on $\theta_0$ for the dominant modes at $k_y \rho_s>0.5$. In a nonlinear simulation for the same surface at an artificially lower $\hat{s}$, where $\gamma^\mathrm{MTM}$ is much more weakly dependent on $\theta_0$, it is found that the (modest) saturated fluxes are largely insensitive to $\mathrm{E\times B}$ shear.

The parameter $\beta_\mathrm{eff}$ from recent theory by Hardman \textit{et al} is useful for describing MTMs in regimes where $q\sim 1$, but as $q$ increases it is not able model the non-monotonic behaviour of $\gamma^\mathrm{MTM}(\theta_0)$ due to the regime being outside the low $\beta$ orderings of the model which was used to exclude $J_\perp$ as $(\nabla \cdot J_\perp) / (\nabla \cdot J_{||}) \propto \beta \big(\frac{qR}{a}\big)^2$. This also has implications for conventional aspect ratio devices as similar behaviour was also found at higher $R/a$. Reactor relevant STs will likely aim to operate with $q_\textrm{min} > 2.0$ \cite{tholerus2024flat, patel2021confinement}, so this inboard destabilisation may be seen. While increasing $\hat{s}$ opens the door to flow shear suppression of the turbulent fluxes, it simultaneously increases the overlap of magnetic islands by reducing the spacing between rational surfaces, and enhances electron heat transport from stochastic fields. Towards the edge of the reactor, $q$ and $\hat{s}$ should be higher than in the core, so $\mathrm{E\times B}$ shear could be more important in this region. Although reactor relevant regimes will be highly self-organised, current profile tailoring can allow for finer control of $q$ and $\hat{s}$ making it a relevant tool in optimising the confinement properties of a future reactor.

\section{Acknowledgements}
The authors would like to thank Walter Guttenfelder for providing the NSTX data, Stuart Henderson and Martin Valovi\v{c} for assisting in generating the MAST data and Jason Parisi for his comments. We would also like to thank Emily Belli and Jeff Candy for assisting with the saturation of MTM simulations performed here. Simulations have been performed on the Marconi National Supercomputing Consortium CINECA (Italy) under the project QLTURB. This work was performed using resources provided by the Cambridge Service for Data Driven Discovery (CSD3) operated by the University of Cambridge Research Computing Service (\url{www.csd3.cam.ac.uk}), provided by Dell EMC and Intel using Tier-2 funding from the Engineering and Physical Sciences Research Council (capital grant EP/T022159/1), and DiRAC funding from the Science and Technology Facilities Council (\url{www.dirac.ac.uk}). This
work was supported by the Engineering and Physical Sciences
Research Council [EP/R034737/1].

\commentout{\appendix

\section{Nonlinear simulations upwinding}
\begin{figure}[!htb]
    \centering
    \includegraphics[width=80mm]{figures/NSTX_upwind_comparison.pdf}
    \caption{Nonlinear electron heat flux for NSTX simulations without $\mathrm{E\times B}$ shear using different upwinding schemes in the parallel direction.}
    \label{fig:upwind_scan_nstx}
\end{figure}
%\section{MAST high ky ETG/PEM}
}
\printbibliography[]

@article{kaye2013NF,
 author={Kaye, S. M. and Gerhardt, S. and Guttenfelder, W. and Maingi, R. and  Bell, R. E. and Diallo, A. and LeBlanc, B.P. and Podesta, M.},
 title={The dependence of H-mode energy confinement and transport on collisionality in NSTX},
 journal={Nuclear Fusion},
 volume={53},
 pages={063005},
 year={2013}
}

@article{tholerus2024flat,
  title={Flat-top plasma operational space of the STEP power plant},
  author={Tholerus, E. and Casson, F. J. and Marsden, S. and Wilson, T. and Brunetti, D. and Fox, P. and Freethy, S. and Hender, T. C. and Henderson, S. S. and Hudoba, A. and others},
  journal={Nuclear Fusion},
  year={2024}
}

@article{hatch2021microtearing,
  title={Microtearing modes as the source of magnetic fluctuations in the JET pedestal},
  author={Hatch, D. R. and Kotschenreuther, M and Mahajan, SM and Pueschel, MJ and Michoski, C and Merlo, G and Hassan, E and Field, AR and Frassinetti, Lorenzo and Giroud, C and others},
  journal={Nuclear Fusion},
  volume={61},
  number={3},
  pages={036015},
  year={2021},
  publisher={IOP Publishing}
}

@article{patel2022linear,
  title={Linear gyrokinetic stability of a high $\beta$ non-inductive spherical tokamak},
  author={Patel, B. S. and Dickinson, David and Roach, CM and Wilson, Howard Read},
  journal={Nuclear Fusion},
  volume={62},
  number={1},
  pages={016009},
  year={2022},
  publisher={IOP Publishing}
}

@article{burrell1997effects,
  title={Effects of E$\times$ B velocity shear and magnetic shear on turbulence and transport in magnetic confinement devices},
  author={Burrell, K. H.},
  journal={Physics of Plasmas},
  volume={4},
  number={5},
  pages={1499--1518},
  year={1997},
  publisher={American Institute of Physics}
}

@article{davies2022kinetic,
  title={Kinetic ballooning modes as a constraint on plasma triangularity in commercial spherical tokamaks},
  author={Davies, R. and Dickinson, D. and Wilson, H.},
  journal={Plasma Physics and Controlled Fusion},
  volume={64},
  number={10},
  pages={105001},
  year={2022},
  publisher={IOP Publishing}
}

@incollection{wilson2020step,
author = {Wilson, H. R. and Chapman, I. T. and Denton, T and Morris, W and Patel, BS and Voss, GM and Waldon, C and the STEP Team},
title = {STEP - on the pathway to fusion commercialization},
booktitle = {Commercialising Fusion Energy},
publisher = {IOP Publishing},
year = {2020},
series = {2053-2563},
type = {Book Chapter},
pages = {8-1 to 8-18},
}

@phdthesis{patel2021confinement,
  title={Confinement physics for a steady state net electric burning spherical tokamak},
  author={Patel, B. S.},
  year={2021},
  school={University of York}
}

@article{guttenfelder2011electromagnetic,
  title={Electromagnetic transport from microtearing mode turbulence},
  author={Guttenfelder, W. and Candy, J. and Kaye, S.M. and Nevins, WM and Wang, E and Bell, RE and Hammett, GW and LeBlanc, BP and Mikkelsen, DR and Yuh, H},
  journal={Physical review letters},
  volume={106},
  number={15},
  pages={155004},
  year={2011},
  publisher={APS}
}

@article{giacomin2023nonlinear,
  title={Nonlinear microtearing modes in MAST and their stochastic layer formation},
  author={Giacomin, M. and Dickinson, D. and Kennedy, D. and Patel, B.S. and Roach, C.M.},
  journal={Plasma Physics and Controlled Fusion},
  volume={65},
  number={9},
  pages={095019},
  year={2023},
  publisher={IOP Publishing}
}

@article{giacomin2024electromagnetic,
  title={On electromagnetic turbulence and transport in STEP},
  author={Giacomin, M. and Kennedy, D. and Casson, F. J. and Ajay, CJ and Dickinson, David and Patel, BS and Roach, CM},
  journal={Plasma Physics and Controlled Fusion},
  volume={66},
  number={5},
  pages={055010},
  year={2024},
  publisher={IOP Publishing}
}

@article{kennedy2023electromagnetic,
  title={Electromagnetic gyrokinetic instabilities in STEP},
  author={Kennedy, D. and Giacomin, M. and Casson, F.J. and Dickinson, D. and Hornsby, W.A. and Patel, B.S. and Roach, C. M.},
  journal={Nuclear Fusion},
  volume={63},
  number={12},
  pages={126061},
  year={2023},
  publisher={IOP Publishing}
}

@article{arbon2020rapidly,
  title={Rapidly-convergent flux-surface shape parameterization},
  author={Arbon, R. and Candy, J. and Belli, E. A.},
  journal={Plasma Physics and Controlled Fusion},
  volume={63},
  number={1},
  pages={012001},
  year={2020},
  publisher={IOP Publishing}
}

@software{Patel_pyrokinetics_2022,
author = {Patel, B. S. and Hill, Peter and Pattinson, Liam and Giacomin, Maurizio and Bokshi, Arkaprava and Kennedy, Daniel and Dudding, Harry G. and Parisi, Jason. F. and Neiser, Tom F. and Jayalekshmi, Ajay C. and Dickinson, David and Ruiz, Juan Ruiz},
doi = {10.21105/joss.05866},
journal = {Journal of Open Source Software},
month = mar,
number = {95},
pages = {5866},
title = {{Pyrokinetics - A Python library to standardise gyrokinetic analysis}},
url = {https://joss.theoj.org/papers/10.21105/joss.05866},
volume = {9},
year = {2024}
}

@article{hallatschek2005giant,
  title={Giant electron tails and passing electron pinch effects in tokamak-core turbulence},
  author={Hallatschek, K. and Dorland, W},
  journal={Physical review letters},
  volume={95},
  number={5},
  pages={055002},
  year={2005},
  publisher={APS}
}

@article{dickinson2013microtearing,
  title={Microtearing modes at the top of the pedestal},
  author={Dickinson, D. and Roach, CM and Saarelma, S and Scannell, R and Kirk, A and Wilson, HR},
  journal={Plasma Physics and Controlled Fusion},
  volume={55},
  number={7},
  pages={074006},
  year={2013},
  publisher={IOP Publishing}
}

@article{miller1998noncircular,
  title={Noncircular, finite aspect ratio, local equilibrium model},
  author={Miller, R. L. and Chu, Ming-Sheng and Greene, JM and Lin-Liu, YR and Waltz, RE},
  journal={Physics of Plasmas},
  volume={5},
  number={4},
  pages={973--978},
  year={1998},
  publisher={American Institute of Physics}
}

@article{candy2018spectral,
  title={Spectral treatment of gyrokinetic shear flow},
  author={Candy, J. and Belli, E. A.},
  journal={Journal of Computational Physics},
  volume={356},
  pages={448--457},
  year={2018},
  publisher={Elsevier}
}

@article{applegate2004microstability,
  title={Microstability in a “MAST-like” high confinement mode spherical tokamak equilibrium},
  author={Applegate, D. J. and Roach, CM and Cowley, SC and Dorland, WD and Joiner, N and Akers, RJ and Conway, NJ and Field, AR and Patel, A and Valovic, M and others},
  journal={Physics of plasmas},
  volume={11},
  number={11},
  pages={5085--5094},
  year={2004},
  publisher={American Institute of Physics}
}

@article{drake1977kinetic,
  title={Kinetic theory of tearing instabilities},
  author={Drake, J. F. and Lee, YC},
  journal={The Physics of Fluids},
  volume={20},
  number={8},
  pages={1341--1353},
  year={1977},
  publisher={American Institute of Physics}
}

@article{hardman2023new,
  title={New linear stability parameter to describe low-$\beta$ electromagnetic microinstabilities driven by passing electrons in axisymmetric toroidal geometry},
  author={Hardman, M. R. and Parra, FI and Patel, Bhavin S and Roach, Colin M and Ruiz, J Ruiz and Barnes, Michael and Dickinson, David and Dorland, W and Parisi, Jason F and St-Onge, D and others},
  journal={Plasma Physics and Controlled Fusion},
  volume={65},
  number={4},
  pages={045011},
  year={2023},
  publisher={IOP Publishing}
}

@article{guttenfelder2012scaling,
  title={Scaling of linear microtearing stability for a high collisionality National Spherical Torus Experiment discharge},
  author={Guttenfelder, W. and Candy, J and Kaye, SM and Nevins, WM and Bell, RE and Hammett, GW and LeBlanc, BP and Yuh, H},
  journal={Physics of Plasmas},
  volume={19},
  number={2},
  pages={022506},
  year={2012},
  publisher={American Institute of Physics}
}

@article{guttenfelder2012simulation,
  title={Simulation of microtearing turbulence in national spherical torus experiment},
  author={Guttenfelder, W. and Candy, J and Kaye, SM and Nevins, WM and Wang, E and Zhang, J and Bell, RE and Crocker, NA and Hammett, GW and LeBlanc, BP and others},
  journal={Physics of Plasmas},
  volume={19},
  number={5},
  pages={056119},
  year={2012},
  publisher={American Institute of Physics}
}

@article{ajay2022,
  title={Microtearing turbulence saturation via electron temperature flattening at low-order rational surfaces},
  author={Ajay, C. J. and McMillan, B. and Pueschel, M. J.},
  journal={arXiv preprint  arXiv:2207.09211v3},
  year={2022}
}

@article{doerk2011gyrokinetic,
  title={Gyrokinetic microtearing turbulence},
  author={Doerk, H. and Jenko, F and Pueschel, MJ and Hatch, DR},
  journal={Physical Review Letters},
  volume={106},
  number={15},
  pages={155003},
  year={2011},
  publisher={APS}
}

@article{hardman2022extended,
  title={Extended electron tails in electrostatic microinstabilities and the nonadiabatic response of passing electrons},
  author={Hardman, M. R. and Parra, FI and Chong, C and Adkins, T and Anastopoulos-Tzanis, MS and Barnes, M and Dickinson, David and Parisi, JF and Wilson, H},
  journal={Plasma Physics and Controlled Fusion},
  volume={64},
  number={5},
  pages={055004},
  year={2022},
  publisher={IOP Publishing}
}

@article{doerk2012gyrokinetic,
  title={Gyrokinetic prediction of microtearing turbulence in standard tokamaks},
  author={Doerk, H. and Jenko, F and G{\"o}rler, T and Told, D and Pueschel, MJ and Hatch, DR},
  journal={Physics of Plasmas},
  volume={19},
  number={5},
  pages={055907},
  year={2012},
  publisher={American Institute of Physics}
}

@article{pueschel2013properties,
  title={Properties of high-$\beta$ microturbulence and the non-zonal transition},
  author={Pueschel, M. J. and Hatch, D. R. and G{\"o}rler, T and Nevins, WM and Jenko, F and Terry, PW and Told, D},
  journal={Physics of Plasmas},
  volume={20},
  number={10},
  pages={102301},
  year={2013},
  publisher={American Institute of Physics}
}

@article{valovic2011collisionality,
  title={Collisionality and safety factor scalings of H-mode energy transport in the MAST spherical tokamak},
  author={Valovi{\v{c}}, M. and Akers, R and De Bock, M and McCone, J and Garzotti, L and Michael, Clive and Naylor, G and Patel, A and Roach, CM and Scannell, R and others},
  journal={Nuclear Fusion},
  volume={51},
  number={7},
  pages={073045},
  year={2011},
  publisher={IOP Publishing}
}

@inproceedings{candy2016crucial,
  title={Crucial role of zonal flows and electromagnetic effects in ITER turbulence simulations near threshold},
  author={Candy, J. and Staebler, GM},
  booktitle={26th IAEA Fusion Energy Conference, Kyoto, Japan, 17--22 October 2016},
  year={2016}
}

@incollection{rechester2020electron,
  title={Electron heat transport in a tokamak with destroyed magnetic surfaces},
  author={Rechester, A. B. and Rosenbluth, M. N.},
  booktitle={Hamiltonian Dynamical Systems},
  pages={684--687},
  year={2020},
  publisher={CRC Press}
}

\end{document}